\newcommand*{\ATLASLATEXPATH}{atlaslatex/latex/}
\PassOptionsToPackage{english}{babel}

\pdfoutput=1

\documentclass[cernpreprint, texlive=2016, UKenglish]{\ATLASLATEXPATH atlasdoc}

\usepackage[subfigure=true]{\ATLASLATEXPATH atlaspackage}

\usepackage{\ATLASLATEXPATH atlasbiblatex}
\usepackage{\ATLASLATEXPATH atlasphysics}
\usepackage{multirow}
\usepackage{siunitx}
\usepackage{arydshln}

\graphicspath{{atlaslatex/logos/}{atlaslatex/figures/}}

\usepackage{EXOT-2016-34-PAPER-defs}

\makeatletter
\DeclareOldFontCommand{\rm}{\normalfont\rmfamily}{\mathrm}
\DeclareOldFontCommand{\sf}{\normalfont\sffamily}{\mathsf}
\DeclareOldFontCommand{\tt}{\normalfont\ttfamily}{\mathtt}
\DeclareOldFontCommand{\bf}{\normalfont\bfseries}{\mathbf}
\DeclareOldFontCommand{\it}{\normalfont\itshape}{\mathit}
\DeclareOldFontCommand{\sl}{\normalfont\slshape}{\@nomath\sl}
\DeclareOldFontCommand{\sc}{\normalfont\scshape}{\@nomath\sc}
\makeatother

\AtlasTitle{Search for a heavy Higgs boson decaying into a $Z$ boson
and another heavy Higgs boson in the $\ell\ell bb$ final state in 
$pp$ collisions at $\sqrt{s}=13$~\TeV\ with the ATLAS detector}

\AtlasAbstract{
A search for a heavy neutral Higgs boson, $A$, decaying into a $Z$ boson
and another heavy Higgs boson, $H$, is performed using
a data sample corresponding to an integrated luminosity of 
\currentluminosity\ from proton--proton collisions at 
$\sqrt{s} = 13$~\TeV\ recorded in 2015 and 2016 by 
the ATLAS detector at the Large Hadron Collider.
The search considers the $Z$ boson decaying to electrons or muons
and the $H$ boson into a pair of $b$-quarks.
No evidence for the production of an $A$ boson is found. Considering each production process separately, 
the 95\% confidence-level upper limits on the $pp\to A\to ZH$ production cross-section times
the branching ratio $H\to bb$ are in the range of 14--830~fb 
for the gluon--gluon fusion process and 26--570~fb for the $b$-associated
process for the mass ranges 130--700~\GeV{} of the $H$ boson and 
230--800~\GeV{} of the $A$ boson. The results
are interpreted in the context of two-Higgs-doublet models.
}

\author{The ATLAS Collaboration}

\AtlasRefCode{EXOT-2016-34}

\PreprintIdNumber{CERN-EP-2018-030}

\AtlasJournal{Phys.\ Lett.\ B.}

\AtlasJournalRef{Phys. Lett. B 783 (2018) 392}
\AtlasDOI{10.1016/j.physletb.2018.07.006}

\hypersetup{pdftitle={ATLAS},pdfauthor={The ATLAS Collaboration}}
\begin{document}

\maketitle


\section{Introduction}
\label{sec:introduction}

After the discovery of a Higgs boson at the Large Hadron Collider
(LHC)~\cite{HIGG-2012-27,CMS-HIG-12-028}, 
one of the most important remaining questions is whether the recently 
discovered particle is part of an extended scalar sector or not. 
Additional Higgs bosons appear in all models with an
extended scalar sector, such as the 
two-Higgs-doublet model (2HDM)~\cite{Lee:1973iz,Branco:2011iw}.
Such extensions are motivated by, and included in, several new physics scenarios, 
such as supersymmetry~\cite{Djouadi:2005gj}, dark 
matter~\cite{Abdallah:2015ter} and axion~\cite{Kim:2008hd} 
models, electroweak baryogenesis~\cite{Cohen:1993nk}  and 
neutrino mass models~\cite{King:2003jb}. 

The addition of a second Higgs doublet leads to five Higgs bosons
after electroweak symmetry breaking. The phenomenology of such a model
is very rich and depends on many parameters, such as the ratio of the
vacuum expectation values of the two Higgs doublets ($\tan\beta$), 
and the Yukawa couplings of the scalar sector~\cite{Branco:2011iw}. 
When CP conservation is assumed, the model contains two CP-even Higgs bosons, $h$ and 
$H$ with $m_{H} > m_{h}$,
one CP-odd, $A$, and two charged scalars, $H^{\pm}$.
There have been many searches for the heavy neutral Higgs bosons 
of the 2HDM at the LHC, including 
$H\to WW/ZZ$ \cite{Aaboud:2017itg,Aaboud:2017fgj,CMS-B2G-16-004,CMS-B2G-16-007}, 
$A/H\to \tau\tau/bb$ \cite{Aaboud:2017sjh,Sirunyan:2018taj,Sirunyan:2018zut}, 
$A\to Zh$ \cite{EXOT-2016-10,CMS-B2G-16-003} 
and $H\to hh$ \cite{HIGG-2013-33,Sirunyan:2017guj}. 
For the interpretation of these searches it is usually assumed that 
the heavy Higgs bosons, $H$ and $A$, are degenerate in mass, i.e.
$m_A = m_H$. 

This assumption of mass degeneracy  is relaxed 
in this Letter by assuming $m_A > m_H$.
Such a case is motivated by electroweak baryogenesis 
scenarios in the context 
of the 2HDM~\cite{Dorsch:2014qja,Turok:1990zg,Fromme:2006cm,Basler:2016obg}. 
For 2HDM electroweak baryogenesis to 
occur, the requirement $m_A > m_H$ is favoured \cite{Dorsch:2014qja} for 
a strong first-order phase transition to take place in the early universe. 
The $A$ boson mass  is also bounded from above to be less than 
approximately $ 800$~\GeV, whereas the lighter CP-even Higgs boson, $h$, 
is required to have properties similar to those of a Standard Model (SM) Higgs boson
and is assumed to be the Higgs boson with mass of $125$~\GeV\ that was discovered 
at the LHC~\cite{Dorsch:2014qja}.
Under such conditions and for large parts of the 2HDM parameter space, the CP-odd Higgs boson, $A$, decays into $ZH$~\cite{Coleppa:2014hxa,Dorsch:2014qja}. 
The production of the $A$ boson in the relevant 2HDM parameter space proceeds
mainly through gluon--gluon fusion and $b$-associated production
at the LHC.

This search for $A\to ZH$ decays uses proton--proton collision data at $\sqrt{s} = 13$~\TeV\  
corresponding to an integrated luminosity of \currentluminosity\ recorded by the ATLAS detector at the LHC. 
The search considers only
 $Z\to \ell\ell$, where $\ell=e,\ \mu$, to take advantage 
of the clean leptonic  final state, and  $H\to bb$, 
because of its large branching ratio.
This final state allows full 
reconstruction of the $A$ boson's decay kinematics. The reconstruction of the
$A$ boson's invariant mass uses the assumed value of the mass of the $H$ boson
to improve its resolution. The final state is also categorised 
by the presence of two or three $b$-tagged jets to take advantage of the $b$-associated production mechanism.
The CMS Collaboration has published a similar search at $\sqrt{s}=8$~TeV~\cite{CMS-HIG-15-001}. 
This Letter reports the result of a search at $\sqrt{s}=13$~TeV, which extends the previous search 
by considering explicitly the gluon--gluon fusion and $b$-associated production processes as well as both 
narrow and wide widths of the $A$ boson. 


\section{ATLAS detector}
\label{sec:atlas}

The ATLAS detector is a general-purpose particle detector, described in detail 
in Ref.~\cite{PERF-2007-01}. It includes an inner detector surrounded by a 
2~T superconducting solenoid, electromagnetic and hadronic
calorimeters and a muon spectrometer with a toroidal magnetic field. 
The inner detector consists of a high-granularity silicon pixel detector, 
including the insertable B-layer~\cite{Capeans:1291633} installed in 2014, a silicon microstrip detector, 
and a straw-tube tracker. It provides precision tracking of charged particles 
with pseudorapidity $|\eta| < 2.5$.\footnote{ ATLAS uses a right-handed coordinate 
system with its origin at the nominal interaction point (IP) 
in the centre of the detector and the $z$-axis along the beam pipe. 
The $x$-axis points from the IP to the centre of the LHC ring, and the
$y$-axis points upward. Cylindrical coordinates ($r$,$\phi$) are used
in the transverse plane, $\phi$ being the azimuthal angle around
the beam pipe. The pseudorapidity is defined
in terms of the polar angle, $\theta$, as $\eta = -\ln \tan(\theta/2)$. 
Transverse momenta are computed from the three-momenta, $\vec{p}$,
 as $\pT = |\vec{p}| \sin\theta$.}  
The calorimeter system covers 
the pseudorapidity range $|\eta| < 4.9$. It is composed of sampling 
calorimeters with either liquid argon or scintillator tiles as the active
medium. The muon spectrometer provides muon identification and 
measurement for $|\eta| < 2.7$. A two-level trigger system~\cite{TRIG-2016-01} is employed 
to select events for offline analysis, which reduced the average 
recorded collision rate to about 1~kHz.

\section{Data and simulation}
\label{sec:samples}

The data used in this search were collected during 2015 and 2016 from $\sqrt{s}=13\, \TeV$ 
proton--proton collisions and
correspond to an integrated luminosity of \currentluminosity, which
includes only data-taking periods where all relevant detector
subsystems were operational. The data sample was collected using 
a set of single-muon and single-electron triggers.
The lowest-\pt\ trigger thresholds depend on the data-taking period
and are in the range of 20--26~\GeV\  for the single-muon triggers and 
24--26~\GeV\ for the single-electron triggers.

Simulated signal events with $A$ bosons produced by gluon--gluon fusion  
were generated 
at leading order with {\MGMCatNLO}~2.3.3 \cite{Alwall:2011uj,Alwall:2014hca} using
{\PYTHIA}~8.210~\cite{Sjostrand:2007gs} with a set of tuned parameters called 
the A14 tune~\cite{ATL-PHYS-PUB-2014-021} for parton showering.
For the generation of $A$ bosons produced in association with $b$-quarks,
{\MGMCatNLO}~2.1.2 \cite{Alwall:2014hca,Frixione:2002ik,Wiesemann:2014ioa} was used following 
Ref.~\cite{deFlorian:2016spz} together with
{\PYTHIA}~8.212 and the A14 tune for parton showering.
The gluon--gluon fusion production used NNPDF2.3LO~\cite{Ball:2012cx} 
as the parton distribution functions~(PDF),
while the $b$-associated production used CT10nlo\_nf4~\cite{Lai:2010vv}.
The signal samples were generated for $A$ bosons with masses in the range of $230$--$800$~\GeV\ 
and widths up to 20\% of the mass and for narrow-width $H$ bosons with masses in the range of $130$--$700$~\GeV.

Background events from the production of $W$ and $Z$ bosons in association with jets were
simulated with {\SHERPA}~2.2.1~\cite{Gleisberg:2008ta} using
the NNPDF3.0NNLO PDF set~\cite{Ball:2014uwa}.
Top-quark-pair production was simulated with 
{\POWHEGBOX}~v2~\cite{Nason:2004rx,Frixione:2007vw,Alioli:2010xd} 
and the CT10nlo PDF set~\cite{Lai:2010vv},
while the electroweak single-top-quark production was simulated
with {\POWHEGBOX}~v1 and the fixed four-flavour PDF set CT10nlo\_f4~\cite{Lai:2010vv}.
The parton shower was performed with {\PYTHIA}~6.428~\cite{Sjostrand:2006za}
using the Perugia 2012 set of tuned parameters~\cite{Skands:2010ak}.
The production of top-quark pairs in association with a vector boson was simulated
using {\MGMCatNLO}~2.2.3 and the NNPDF3.0NLO PDF set, whereas 
{\PYTHIA}~8.186 was used for the parton shower with the A14 tune.
Production of $WW$, $ZZ$ and $WZ$ pairs was simulated using {\SHERPA}~2.2.1
and the NNPDF3.0NNLO PDF set. Finally, SM Higgs boson
production in association with a $Z$ boson was generated
with {\POWHEGBOX}~v2 and the NNPDF3.0NLO PDF set, whereas
the parton shower was performed with {\PYTHIA}~8.186 using the AZNLO tune~\cite{Aad:2014xaa}.

The modelling of bottom- and charm-hadron decays was performed with the 
EvtGen~v1.2.0 package~\cite{Lange:2001uf} for all samples apart from 
those simulated with {\SHERPA}.
The simulated events were overlaid with 
inelastic proton--proton collisions to account for the
effect of multiple interactions occurring in the same
and neighbouring bunch crossings (`pile-up').
These events were generated using {\PYTHIA}~8 with the A2
tune~\cite{ATL-PHYS-PUB-2012-003} and the MSTW2008LO 
PDF set~\cite{Martin:2009iq}.
The events were reweighted so that the distribution of the average
number of interactions per bunch crossing agreed with the data.

All generated background samples were passed through the
{\GEANT}4-based~\cite{Agostinelli:2002hh}  
detector simulation~\cite{SOFT-2010-01} of the ATLAS detector.  
The ATLFAST2 simulation~\cite{SOFT-2010-01} 
was used for the signal samples to allow for the generation of many different
$A$ and $H$ boson masses.
The simulated events were reconstructed in the same way as the data.


\section{Object reconstruction}
\label{sec:objects}

Electrons are reconstructed from energy clusters in
the electromagnetic calorimeter that are matched
to tracks in the inner detector~\cite{PERF-2013-05}. 
Electrons are required to have $|\eta| < 2.47$ and $\pt > 7~\GeV$. 
To distinguish electrons from jets, isolation and quality requirements are applied~\cite{ATLAS-CONF-2016-024}.
The isolation requirements (the `LooseTrackOnly' working point) are defined by
the $\pT$ of tracks within cones around the
electron with a size that decreases as a function of the transverse energy.
The quality requirements (the `Loose' working point) refer to
both the inner detector track and the calorimeter shower shape.
The efficiency for an electron to be reconstructed and meet these criteria 
is about 85\%  for electron $\pt > 7~\GeV$ and increases to about 90\% for $\pt >27~\GeV$. 

Muons are reconstructed by matching tracks reconstructed 
in the inner detector to tracks or track
segments in the muon spectrometer~\cite{PERF-2015-10}.
Muons used for this search must have $|\eta| < 2.5$ and $\pt > 7~\GeV$, 
and are required to satisfy `LooseTrackOnly' isolation requirements, similar to those 
used for electrons, as well as inner detector and muon 
spectrometer track `Loose' quality criteria, corresponding to an
efficiency of about 97\%.

Jets are reconstructed using the anti-$k_t$ 
algorithm~\cite{Cacciari:2008gp,Cacciari:2011ma} with radius 
parameter $R = 0.4$ from clusters of energy deposits
in the calorimeter system~\cite{PERF-2014-07}. 
Candidate jets are required to have 
$\pt > 20~\GeV$ ($\pt > 30~\GeV$) for $|\eta| < 2.5$ 
$(2.5 < |\eta| < 4.5)$. Low-\pt\ jets from pile-up are rejected
by a multivariate algorithm that uses properties of
 the reconstructed tracks in the event~\cite{PERF-2014-03}.

Jets containing $b$-hadrons 
are selected using a multivariate tagging algorithm ($b$-tagging)~\cite{PERF-2012-04,ATL-PHYS-PUB-2016-012}. 
The energy of the tagged jet ($b$-jet) is corrected for the average energy loss 
from semileptonic decays of $b$-hadrons and out-of-jet-cone tracks with
large impact parameters~\cite{Aaboud:2017xsd}.
The $b$-tagging efficiency for the jet \pt{} range used in this
analysis is between 65\% and 75\%. Applying the $b$-tagging algorithm
reduces the number of light-flavour ($c$-quark) jets by a factor
of 250--550 (10--20), depending on the jet kinematics.

When electrons, muons and jets are spatially close, these algorithms can lead 
to ambiguous identifications. An overlap removal procedure~\cite{Aaboud:2017xsd} 
is therefore applied to uniquely identify these objects. 

The missing transverse momentum, \met, is computed
using reconstructed and calibrated leptons, photons and jets~\cite{ATL-PHYS-PUB-2015-027}.
Tracks from the primary vertex\footnote{The primary vertex is taken to be the 
reconstructed vertex with the highest $\Sigma p_{\mathrm{T}}^2$ 
of the associated tracks.} which are not associated with any identified lepton
or jet are
also taken into account in the \met{} reconstruction~\cite{PERF-2014-04}.


\section{Event selection}
\label{sec:selection}

The decay $A\to ZH\to \ell\ell bb $  features a pair of
oppositely charged, same flavour leptons and two $b$-jets. Three resonances
can be formed by combining the selected objects: the
$Z$ boson ($\ell\ell$), the $H$ boson ($bb$) and the $A$ boson
($\ell\ell bb$). Moreover, additional $b$-jets 
may be present if the $A$ boson is produced via the $b$-associated
production mechanism. These features are used to define
the event selection as summarised in Table~\ref{tab:selections}.

The events recorded by the single-muon and the single-electron
triggers are required to contain exactly two muons or two electrons, respectively.
At least one of the leptons must have $\pt>27$~GeV.
Only events that contain a primary vertex  with
at least two associated tracks with $\pt > 400~\MeV$~\cite{PERF-2015-01} 
are considered.
In the case of muons, they are 
required to have opposite electric charges. No such requirement is applied
to electrons due to their non-negligible charge misidentification rates resulting from conversions of bremsstrahlung photons.
The invariant mass of the lepton pair, $m_{\ell\ell}$, must be in 
the range of 80--100~\GeV\ to be compatible with the mass of the $Z$ boson.

The $H\to bb$ decay is reconstructed by requiring  at
least two $b$-jets with the highest-\pt\ one having
$\pt > 45~\GeV$. When more than two $b$-jets are present,
the two highest-\pt\ $b$-jets are considered to be from the $H$ decay.
Requiring $b$-jets increases the fraction of top-quark background
in the signal region, including top-quark pair and
single-top-quark production. This is reduced by requiring
$\met /\sqrt{H_{\rm T}} < 3.5~\GeV^{1/2}$, where $H_{\rm T}$ is the scalar sum of
the \pt\ of all jets and leptons in the event.
In addition, a requirement that reduces the $Z$+jets background is
also applied: $\sqrt{\Sigma \pT^2}/m_{\ell\ell bb} > 0.4$, where
$m_{\ell\ell bb}$ is the four-body invariant mass of the two-lepton, 
 two-$b$-jet system assigned to the $A$ boson
and the summation is performed over the \pt\ of these objects.

Subsequently, two categories are defined: the 
$n_b=2$ category, which contains events with exactly two $b$-jets, and the
$n_b \geq 3$ category, which contains events with three  or more $b$-jets. 
For the gluon--gluon fusion production, 94\%--97\% of the events passing the 
above selection fall into the $n_b=2$ category, depending on the assumed $m_A$ and $m_H$.
However for the $b$-associated production, 27\%--36\% fall into the $n_b \geq 3$ category.
The remaining $b$-associated produced signal events are categorised as $n_b=2$ events,
even though more than two $b$-jets are expected, due to the relatively soft \pt\ spectrum 
of the associated $b$-jets and the geometric acceptance of the tracker.

Finally, the invariant mass of the two leading 
$b$-jets, $m_{bb}$, must be compatible with the assumed $H$ boson mass 
by satisfying the requirement of 
$ 0.85\cdot m_H - 20~\GeV < m_{bb} < m_H + 20~\GeV  $ for the
$n_b=2$ category, and
$ 0.85\cdot m_H - 25~\GeV < m_{bb} < m_H + 50~\GeV  $ for the
$n_b \geq 3$ category.
The wider window for $n_b\geq 3$ is motivated by a slightly degraded resolution
due to potential $b$-jet mis-assignments (see later).
The overall signal efficiency of the $n_b=2$ category after this 
requirement is 5\%--11\% (3\%--7\%) for gluon--gluon fusion
($b$-associated production), depending
on the $m_A$ and $m_H$ values. Similarly, the efficiency of the $n_b \geq 3$ category
is 2\%--4\% for the $b$-associated production.
The signal region selection is summarised in Table~\ref{tab:selections}.

\begin{table}[h!]
\caption{Summary of the event selection for signal and control regions.}
\centering
\small
\setlength\tabcolsep{5pt} 
\begin{tabular}{l|lll}
\hline
\multicolumn{4}{c}{Single-electron or single-muon trigger} \\
\multicolumn{4}{c}{Exactly 2 leptons ($e$ or $\mu$) ($\pt > 7$~\GeV) 
with the leading one having $\pt > 27$~\GeV} \\
\multicolumn{4}{c}{Opposite electric charge for $\mu\mu$ or $e\mu$ pairs;~~ $80\ \GeV < m_{\ell\ell},\ m_{e\mu} < 100\ \GeV$, $\ell = e,\ \mu$} \\
\multicolumn{4}{c}{At least 2 $b$-jets ($\pt > 20\, \GeV$) with one of them having $\pt > 45\, \GeV$} \\
\multicolumn{4}{c}{$\met /\sqrt{H_{\rm T}} < 3.5~\GeV^{1/2}$, $\sqrt{\Sigma \pT^2}/m_{\ell\ell bb} > 0.4$} \\
\hline \\[-3mm]
  &    $n_b =2$ category   &\hspace*{0.5cm}&  $n_b \ge 3$ category \\ 
\hline \\[-3mm]
              & Exactly 2 $b$-tagged jets    && At least 3 $b$-tagged jets \\
\hdashline 
\\[-3mm]
Signal & $ee$ or $\mu\mu$ pair && $ee$ or $\mu\mu$ pair \\
region   & $ 0.85\cdot m_H\! -\! 20~\GeV\! < m_{bb}\! < m_H\! +\! 20~\GeV $ && $ 0.85\cdot m_H\! -\! 25~\GeV\! < m_{bb}\! < m_H\! +\! 50~\GeV$ \\
\hdashline 
\\[-3mm]
Top & $e\mu$  pair && $e\mu$  pair \\
control region  & $ 0.85\cdot m_H\! -\! 20~\GeV\! < m_{bb}\! < m_H\! +\! 20~\GeV$ && $ 0.85\cdot m_H\! -\! 25~\GeV\! < m_{bb}\! < m_H\! +\! 50~\GeV$ \\
\hdashline 
\\[-3mm]
$Z+$jets & $ee$ or $\mu\mu$ pair && $ee$ or $\mu\mu$ pair \\
control region   & $m_{bb} < 0.85\cdot m_H - 20~\GeV$  && $ m_{bb} < 0.85\cdot m_H - 25~\GeV $ \\
                  & or $ m_{bb} > m_H + 20~\GeV$  &&  or $ m_{bb} > m_H + 50~\GeV$ \\
 \hline
\end{tabular}
\label{tab:selections}
\end{table}

The $m_{\ell\ell bb}$ distribution after the $m_{bb}$ requirement 
is used to discriminate between signal and background. 
To improve the $\mllbb$  resolution, the 
$bb$ system's four-momentum components are scaled to match the
assumed $H$ boson mass and the $\ell\ell$ system's four-momentum
components are scaled to match the $Z$ boson mass. This procedure, performed
after the event selection, 
improves the $m_{\ell\ell bb}$ resolution by a factor 
of two without significantly distorting the background distributions,
resulting in an $A$ boson mass resolution of 0.3\%--4\%. 

The dominant backgrounds after these selections are from
$Z$+jets and top-quark production. For top-quark-pair production, a very pure ($>99\%$ of predicted events) control
region is used to determine
the normalisation of the background, whereas its shape in the signal region
is taken from the simulation. This control region is defined
by keeping the same selection as discussed previously, 
apart from an opposite-flavour lepton criterion, i.e., 
an opposite-charge $e\mu$ pair is required instead of an $ee$ or $\mu\mu$ pair
(see also Table~\ref{tab:selections}). 
The shape of the $Z$+jets background distribution is obtained
from  simulation and the normalisation is extracted from data together 
with the signal (see also Section~\ref{sec:fit}). 
This procedure is possible because of the very 
different shapes of the $m_{\ell\ell bb}$ distributions from signal and $Z$+jets 
events. The normalisation of the  $Z$+jets production is further
constrained by a control region defined by inverting the $m_{bb}$
window criterion for each $H$ boson mass hypothesis 
(see also Table~\ref{tab:selections}). 
The control regions are distinct for the $n_b = 2$ and the $n_b \geq 3$
categories, since the accuracy of the background simulation 
depends on the number of $b$-jets present in the event.
Backgrounds from diboson, single top, and Higgs boson production, as well as
top-quark-pair production in association with a vector boson, give a
typical contribution of $\sim 5$\% to the total background. Their shapes
are taken from simulation, whereas they are normalised using
precise inclusive cross-sections calculated from theory. 
The diboson samples are normalised using next-to-next-to-leading-order (NNLO) cross-sections~\cite{Gehrmann:2014fva,Grazzini:2016swo,Grazzini:2015hta,Cascioli:2014yka}. 
Single-top-quark production and top-quark-pair production in
association with vector bosons are normalised to next-to-leading-order (NLO) cross-sections from
Refs.~\cite{Kidonakis:2010tc,Kidonakis:2011wy,Kidonakis:2010ux} and 
Ref.~\cite{Alwall:2014hca}, respectively. 
The normalisation of the Higgs boson production in association with 
a vector boson follows the recommendations of Ref.~\cite{deFlorian:2016spz}
using NNLO QCD and NLO electroweak corrections.


\section{Signal modelling}
\label{sec:signal}

The good $\mllbb$ mass resolution together with the fact that 
theory models often predict $A$ bosons with large widths inflates the number of signal mass and width
hypotheses that need to be considered. For this reason, the $\mllbb$ distributions
are taken from simulation of a limited number of $(m_A,m_H)$ mass points and
an interpolation using analytical functions is employed for the rest.

\begin{figure}[hb]
\centering
\subfigure[]{\includegraphics[width=0.49\textwidth]{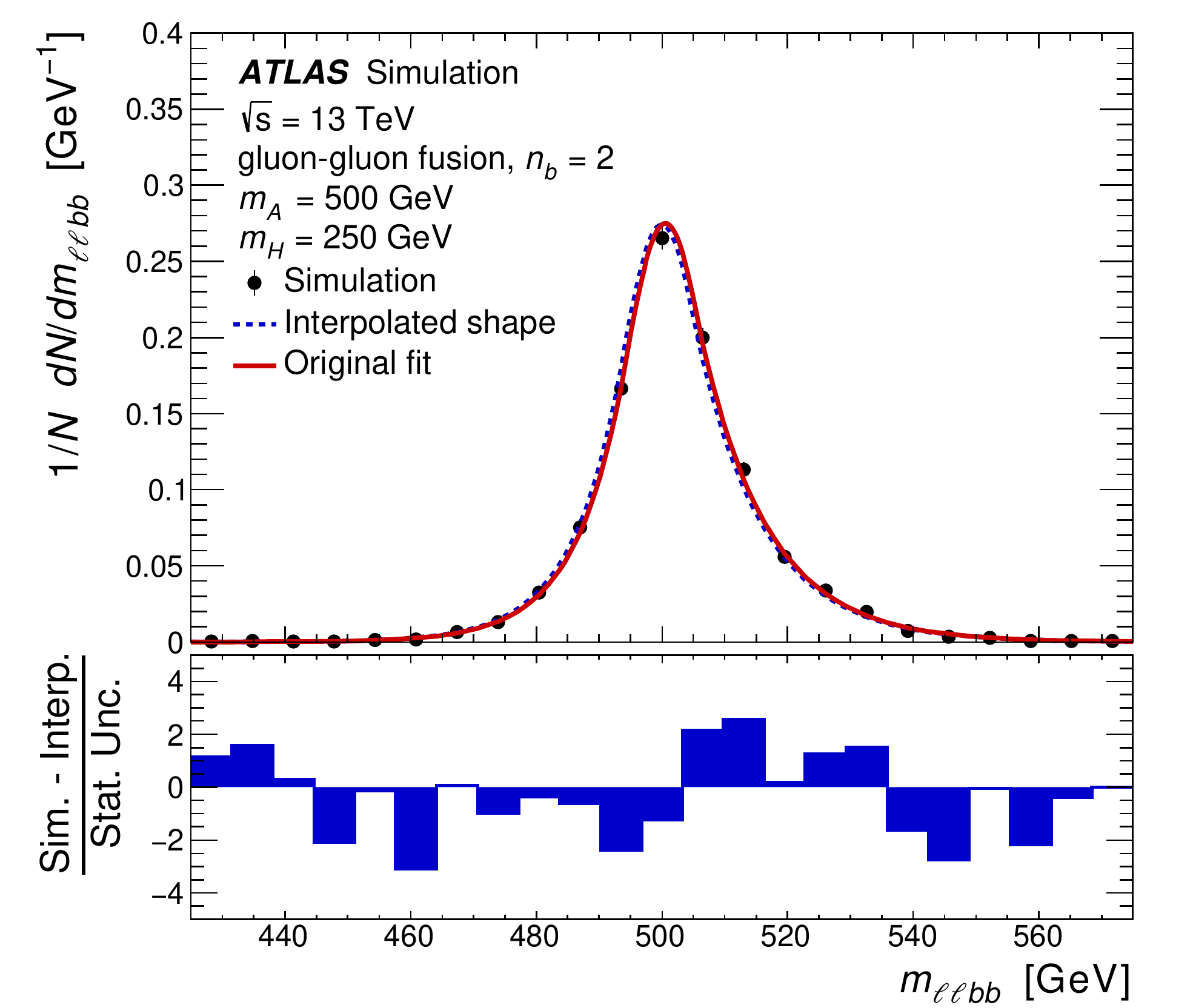}} 
\subfigure[]{\includegraphics[width=0.49\textwidth]{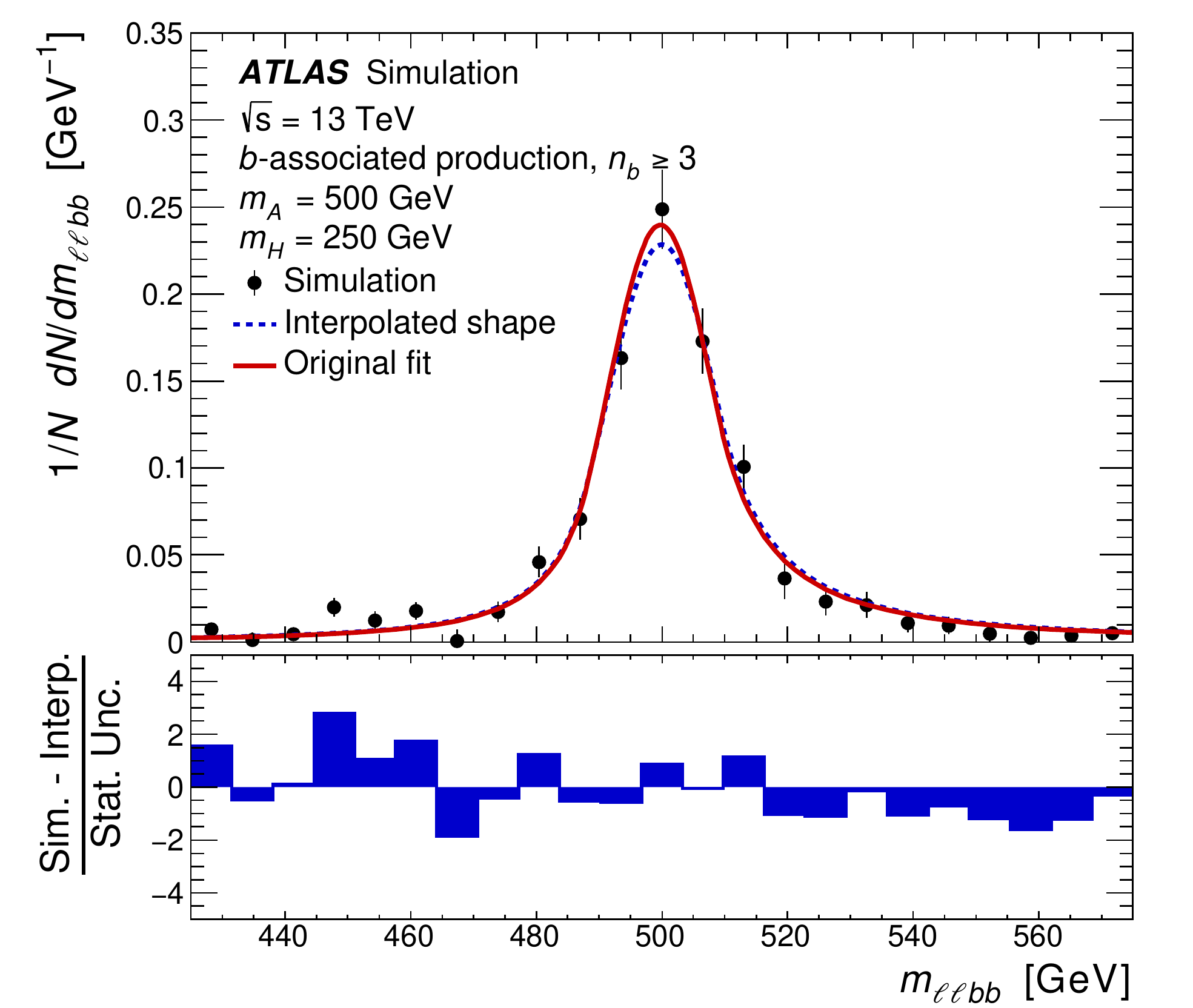}}
\caption{Simulated signal $\mllbb$ distributions (closed circles) 
assuming  $m_A=500$~\GeV\ and $m_H=250$~\GeV\ for the following cases:
(a) the gluon--gluon fusion in the $n_b=2$ category and 
(b) $b$-associated production in the $n_b \geq 3$ category. 
Signal parameterisations are overlaid for comparison. The solid curves are from  parameter values obtained directly from the fits to the simulated distributions, whereas the dashed curves use the interpolated parameter values. The differences between the simulation and the interpolated shape divided by the statistical uncertainties of the simulation are shown in the bottom panels. The distributions for the $n_b=2$ category of the $b$-associated production are similar to the $n_b\ge 3$ shape shown.}
\label{fig:signal_model}
\end{figure}

The  $\mllbb$ distributions for $A$ bosons produced by gluon--gluon fusion and with
negligible widths compared with the experimental resolution 
are found to be adequately described by the {\em ExpGaussExp}~(EGE) 
function~\cite{Das:2016stf}: 
\begin{displaymath}\label{eq:EGE}
  \begin{array}{rclcl}
   f_{\mathrm{EGE}}(m;a,\sigma,k_L,k_H) & = & 
         \left\{\begin{array}{l}  
             {\mathrm{e}}^{\frac{1}{2}k_L^2+k_L\left(\frac{m-a}{\sigma}\right)} \\ \\[-3.5mm]
             {\mathrm{e}}^{-\frac{1}{2}\left(\frac{m-a}{\sigma}\right)^2} \\ \\[-3.5mm]
             {\mathrm{e}}^{\frac{1}{2}k_H^2-k_H\left(\frac{m-a}{\sigma}\right)} 
         \end{array}\right.
         & {\rm for} & 
         \left\{\begin{array}{l}
             \frac{m-a}{\sigma}\le - k_L  \\ \\[-3.5mm]
             -k_L <\frac{m-a}{\sigma}\le k_H  \\ \\[-3.5mm]
             \frac{m-a}{\sigma} > k_H
         \end{array}\right.
  \end{array}
\end{displaymath}

On the other hand, $\mllbb$ distributions for $A$ bosons from $b$-associated production, also with 
negligible widths compared with the experimental resolution, 
are better described by a double-Gaussian Crystal Ball (DSCB) function~\cite{Oreglia:1980cs}:
\begin{displaymath}\label{eq:DGCB}
  \begin{array}{rclcl}
   f_{\mathrm{DSCB}}(m;a,\sigma,k_L,k_H,n_1,n_2) & = & 
         \left\{\begin{array}{l}  
             g(m;a,-\sigma,k_L,n_1)\cdot  {\mathrm{e}}^{-\frac{1}{2}k_L^2} \\ \\[-3.5mm]
             {\mathrm{e}}^{-\frac{1}{2}\left(\frac{m-a}{\sigma}\right)^2} \\ \\[-3.5mm]
             g(m;a,\sigma,k_H,n_2)\cdot {\mathrm{e}}^{\frac{1}{2}k_H^2}
         \end{array}\right.
         & {\rm for} & 
         \left\{\begin{array}{l}
             \frac{m-a}{\sigma}\le - k_L  \\ \\[-3.5mm]
             -k_L <\frac{m-a}{\sigma}\le k_H  \\ \\[-3.5mm]
             \frac{m-a}{\sigma} > k_H
         \end{array}\right.
  \end{array}
\end{displaymath}
where 
$g(m;a,\sigma,k,n) = [ (|k|/n)(n/|k| - |k| + (m-a)/\sigma) ]^{-n}  $.
Both functions consist of a Gaussian core with mean $a$ and variance
$\sigma^2$, whereas the rest of the parameters ($k_L$, $k_H$, $n_1$, $n_2$)
describe the tails. The DSCB function describes better than the EGEE function
 the slightly longer tails of the mass distribution
of the $b$-associated production compared to gluon--gluon fusion. This
is due to the few cases in which the $b$-quark produced in association with the 
Higgs boson is taken to be one of the $b$-quarks from the Higgs boson decay. 
The values of the function parameters are extracted from unbinned maximum-likelihood
fits to the simulated $\mllbb$ distributions. The core mean, $a$, is 
parameterised using a linear function of $m_A$. The core width, $\sigma$, is 
observed to monotonically increase with $\Delta m\equiv m_A-m_H$ and is 
parameterised with a third-degree polynomial. The rest of the parameters 
are largely constant and are fixed to their average values from the fits, with
the exception of mass points with $\Delta m = 100~\GeV$. The distributions
at mass points with $\Delta m = 100~\GeV$ correspond to the smallest
mass splitting considered in this search and are close to the kinematic cutoff.
Their non-core parameters are fixed to the average fit values obtained from 
signal samples with this mass splitting only. As an example of the performance
of this procedure, Figure~\ref{fig:signal_model} shows a comparison for 
the $(m_A,m_H)=(500,250)$~\GeV\ mass point between the simulated distributions 
and the parametric functions described previously. The cores of the $\mllbb$ distributions 
are well-parameterised by the chosen functional forms. 
The small differences seen in the tails of some distributions  
between the functional forms and the simulations have only negligible 
effects on the final results, and moreover they are included as a
source of systematic uncertainty.

\begin{figure}[hb]
\centering
\subfigure[]{\includegraphics[width=0.49\textwidth]{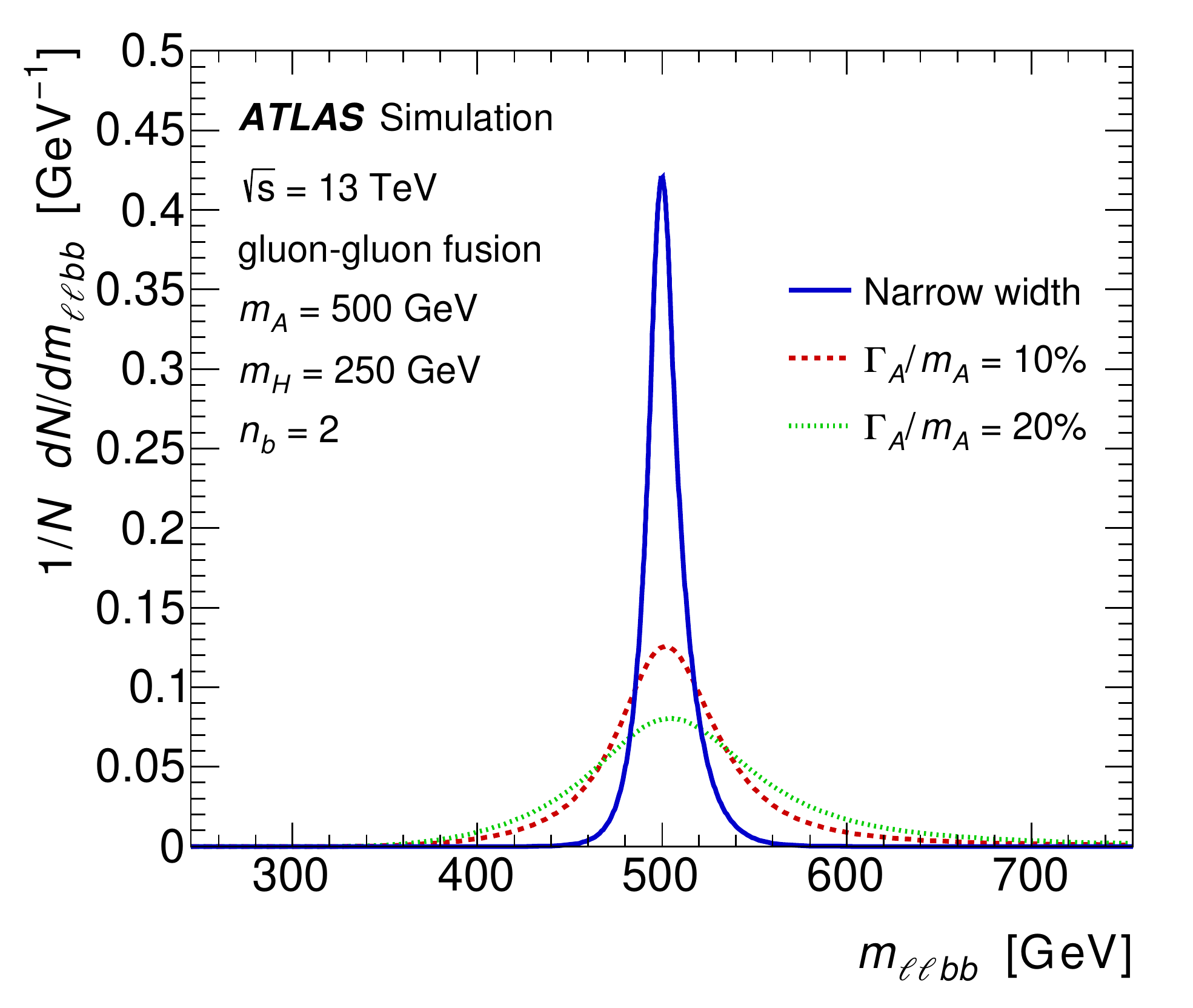}}
\subfigure[]{\includegraphics[width=0.49\textwidth]{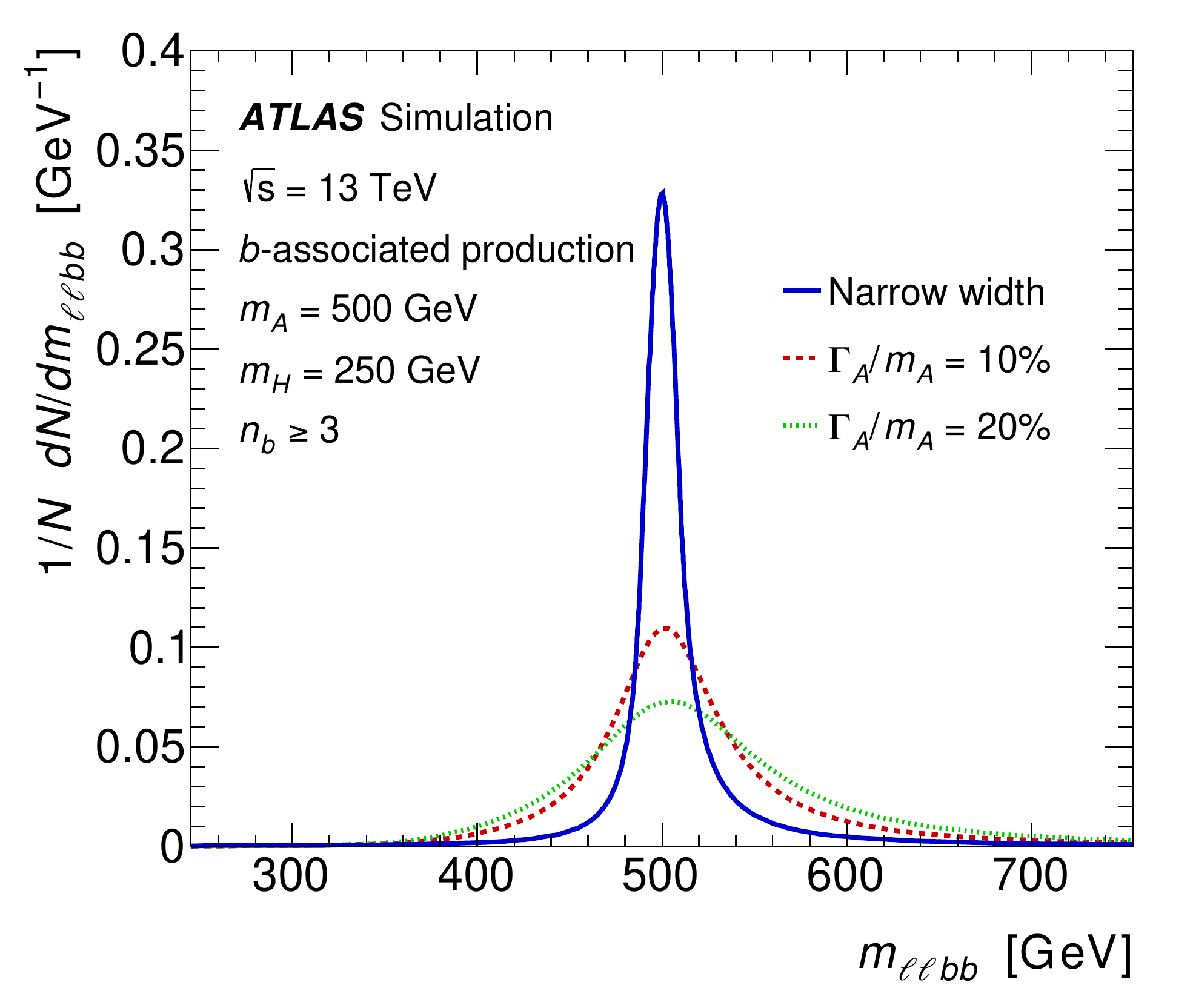}}
\caption{The interpolated signal $\mllbb$ distribution shapes assuming
$m_A=500$~\GeV\ and $m_H=250$~\GeV and various $A$ boson widths for
the following cases:
(a) gluon--gluon fusion in the $n_b=2$ category and 
(b) $b$-associated production in the $n_b \geq 3$ category. The distributions for the $n_b=2$ category of the $b$-associated production are similar to the $n_b\ge 3$ shape shown.}
\label{fig:signal_model-lw}
\end{figure}

The previously described parameterisation applies to signal samples 
generated with narrow-width $A$ bosons. In some regions of 2HDM parameter 
space relevant to this analysis, 
the $A$ boson's width is significant compared with the detector resolution while the $H$ boson's width
remains negligible.
 In order to model the $\mllbb$ shape of
$A$ bosons with large natural widths, a modified Breit--Wigner 
distribution\footnote{The
modification is the multiplication of the Breit--Wigner distribution with a
log-normal distribution to account for the distortion due to the event 
selection.} is convolved with the EGE and DSCB functions. The procedure is validated
by comparing the results of the convolution with those of the simulated samples of $A$ bosons with
large natural widths. Widths of up to 20\% of the $A$ boson mass are 
considered. An example of signal distributions with large natural
widths is shown in Figure~\ref{fig:signal_model-lw} for the same signal
points used in Figure~\ref{fig:signal_model}.

Finally, the signal efficiencies for the interpolated mass points are obtained through separate 
two-dimensional interpolations on the $(m_A,m_H)$ plane using
thin plate splines~\cite{Duchon1976}.


\section{Fit model and systematic uncertainties}
\label{sec:fit}

The $\mllbb$ distribution is expected to exhibit a resonant structure if signal events are present, while background events result in a smooth shape. Therefore $\mllbb$ is chosen as the final signal and background discriminating variable. The shape differences in the $\mllbb$ distribution between the signal and background contributions are exploited through binned maximum-likelihood fits of the signal-plus-background hypotheses to extract potential signal contributions. The fits are based on the statistical framework described in Refs.~\cite{Moneta:2010pm,RooFit,Cowan:2010js}. For a given mass hypothesis of $(m_A,m_H)$, the likelihood is constructed as the product of Poisson statistics in $\mllbb$ bins:
\[L(\left.\mu,\vec{\alpha},\vec{\theta}\right|m_A,m_H) = \prod_{i={\rm bins}} {\rm Poisson}\left(N_i\left|\left(\mu\times S_i(m_A,m_H,\vec{\theta})+B_i(\vec{\alpha},\vec{\theta})\right)\right.\right)\cdot G(\vec{\theta}).\]
Here $N_i$ is the number of observed events and $S_i(m_A,m_H,\vec{\theta})$ and $B_i(\vec{\alpha},\vec{\theta})$ are the expected number of signal and estimated background events in bin $i$. The vector $\vec{\alpha}$ represents free background normalisation scale factors (described later) and vector $\vec{\theta}$ denotes all non-explicitly listed parameters of the likelihood function such as nuisance parameters associated with systematic uncertainties. The function $G(\vec{\theta})$ represents constraints on $\vec{\theta}$. The parameter of interest, $\mu$, is a multiplicative factor to the expected signal rate and is called the signal-strength parameter. The $\mllbb$ bin widths are chosen according to the expected detector resolution and taking into account the statistical uncertainty related to the number of background Monte Carlo events. The bin centres are adjusted such that at least 68\% of the test signal is contained in one bin. 
Only the $n_b=2$ category is considered for the gluon--gluon fusion production while both the $n_b=2$ and $n_b\ge 3$ categories are included in the likelihood calculation for the $b$-associated production.

For each bin, $S_i$ is calculated from the total integrated luminosity, the theoretical cross-section for the signal and its selection efficiency. The sum of all background contributions in the bin, $B_i$, is estimated from simulation. However, the $t\bar{t}$ and $Z$+jets control regions are included in the likelihood calculation as one bin each, to help constrain their respective contributions in the signal regions.  This is achieved by introducing two free normalisation scale factors, represented by $\vec{\alpha}$, for each category for the two most relevant background contributions:  one for $t\bar{t}$ and the other for the heavy-flavour component ($Z$+h.f.) of the $Z$+jets contribution. These scale factors are applied to their respective contributions estimated from the simulations and their values are determined from the fit. Typical values of scale factors are close to unity. Taking $(m_A,m_H)=(700,200)$~\GeV{} as an example, the $Z$+h.f. scale factor is $1.12\pm 0.09$ for the $n_b=2$ category and $1.1\pm 0.2$ for the $n_b\ge 3$ category. Similarly, the $t\bar{t}$ scale factors are $0.96\pm 0.06$  and $1.2\pm 0.2$ for the two corresponding categories. 

Systematic uncertainties are incorporated in the likelihood as nuisance parameters with either Gaussian or log-normal constraint terms. They include both the experimental and theoretical sources of uncertainty. Experimental uncertainties comprise those in the luminosity measurement, trigger, 
object identification, energy/momentum scale and resolution as well as underlying event and pile-up modelling. These uncertainties, discussed in Refs.~\cite{Aaboud:2017xsd,Aaboud:2017itg}, impact the simulations of signal and background processes. Theoretical uncertainties include both the signal and background modelling. For the signal modelling, uncertainties due to the factorisation and renormalisation  scale choice, the initial- and final-state radiation treatment and the PDF choice are considered. Additional systematic uncertainties are assigned to cover the differences in signal efficiencies and $\mllbb$ parameter values between the interpolations and the simulations.  
For the background modelling, the most important sources of systematic uncertainty are from the modelling of the $m_{bb}$ and the $p_{\mathrm{T}}^{\ell\ell}$ distributions of $Z$+jets. They are taken to be the difference between the data and simulation of the selected samples before the event categorisation and the $m_{bb}$ requirement. The samples are dominated by the $Z$+jets contribution, and any potential signal contamination is expected to be negligible. For other background processes, they are obtained by varying the factorisation and renormalisation scales, the amount of initial- and final-state radiation, and the choices of PDF parameterisations. 

The effect of these systematic uncertainties on the search is studied using the signal-strength parameter $\mu$ for hypothesised signal production. Uncertainties having the largest impact depend on the choice of $(m_A, m_H)$ signal point. Table~\ref{tab:systematics} shows the relative uncertainties in the best-fit $\mu$ value from the leading sources of systematic uncertainty for two example mass points of both gluon--gluon fusion and $b$-associated production of a narrow-width $A$ boson. The leading sources of systematic uncertainty are similar for other mass points studied and for larger $A$ boson widths. For all cases, the limited size of the simulated samples has the largest impact on the search sensitivity among all the sources of systematic uncertainty. While systematic uncertainties and the statistical uncertainty of the data have comparable impact at low masses, the search sensitivity is mostly determined at high masses by the limited size of the data sample.  

\begin{table}[h!]
\caption{The effect of the most important sources of uncertainty on the signal-strength parameter at two example mass points of $(m_A,m_H)=(230,130)$~\GeV{} and $(m_A,m_H)=(700,200)$~\GeV{} for both the gluon--gluon fusion and $b$-associated production of a narrow-width $A$ boson. The signal cross-sections are taken to be the expected median upper limits (see Section~\ref{sec:results}). 
JES and JER stand for jet energy scale and jet energy resolution, `Sim. stat.' for simulation statistics, and `Bkg. model.' for the background modelling.}
\centering
\small
\setlength\tabcolsep{5pt} 
\begin{tabular}{lSlSclSlS}
\hline
\multicolumn{4}{c}{Gluon--gluon fusion production}          && \multicolumn{4}{c}{$b$-associated production} \\ 
\multicolumn{2}{c}{$(230,\,130)$\,\GeV} &\multicolumn{2}{c}{$(700,\,200)$\,\GeV}  && \multicolumn{2}{c}{$(230,\,130)$\,\GeV} & \multicolumn{2}{c}{$(700,\,200)$\,\GeV} \\ \cline{1-4} \cline{6-9}
{Source}   & {$\Delta\mu/\mu$ [\%]} & {Source}  & {$\Delta\mu/\mu$ [\%]} && {Source}  & {$\Delta\mu/\mu$ [\%]}  & {Source}  & {$\Delta\mu/\mu$ [\%]} \\ \hline 
Data stat.     &    32  & Data stat.    &  49   &&  Data stat.    &  35  & Data stat.      & 46    \\
Total syst.    &    36  & Total syst.   &  22   &&  Total syst.   &  38  & Total syst.     & 26    \\ \hline
Sim. stat.     &    22  & Sim. stat.    &  10   &&  Sim. stat.    &  26  & Sim. stat.      & 12    \\
Bkg. model.    &    16  & Bkg. model.   &  10   &&  $b$-tagging   &  14  & Bkg. model.     & 11    \\
JES/JER        &    12  & Theory        & 9.1   &&  JES/JER       &  11  & $b$-tagging     & 10    \\
$b$-tagging    &   9.9  & $b$-tagging   & 8.5   &&  Bkg. model.   & 9.8  & Theory          & 6.8   \\
Theory         &   7.5  & Leptons       & 4.2   &&  Theory        & 7.0  & JES/JER         & 6.2   \\ \hline
\end{tabular}
\label{tab:systematics}
\end{table}


\section{Results}
\label{sec:results}

\def \minGGAlim {14}
\def \maxGGAlim {830}
\def \minGGAlimExp{24.1}
\def \maxGGAlimExp{469}
\def \minGGAlimPoint{(800, 140)}
\def \maxGGAlimPoint{(240, 130)}

\def \minBBAlim {26}
\def \maxBBAlim {570}
\def \minBBAlimExp{46}
\def \maxBBAlimExp{360}
\def \minBBAlimPoint{(780, 680)}
\def \maxBBAlimPoint{(240, 130)}

The $\mllbb$ distributions from different $m_{bb}$ mass windows are scanned for potential
excesses beyond the background expectations through signal-plus-background fits. 
The scan is performed in steps of 
10~\GeV{} for both the $m_A$ range 230--800~\GeV{} and the $m_H$ range 130--700~\GeV, such that $m_A-m_H \geq 100\ \GeV$. The step sizes are chosen to be 
compatible with the detector resolution for $\mllbb$ and $m_{bb}$. 

\begin{figure}[htb]
\centering
\subfigure[]{\includegraphics[width=0.49\textwidth]{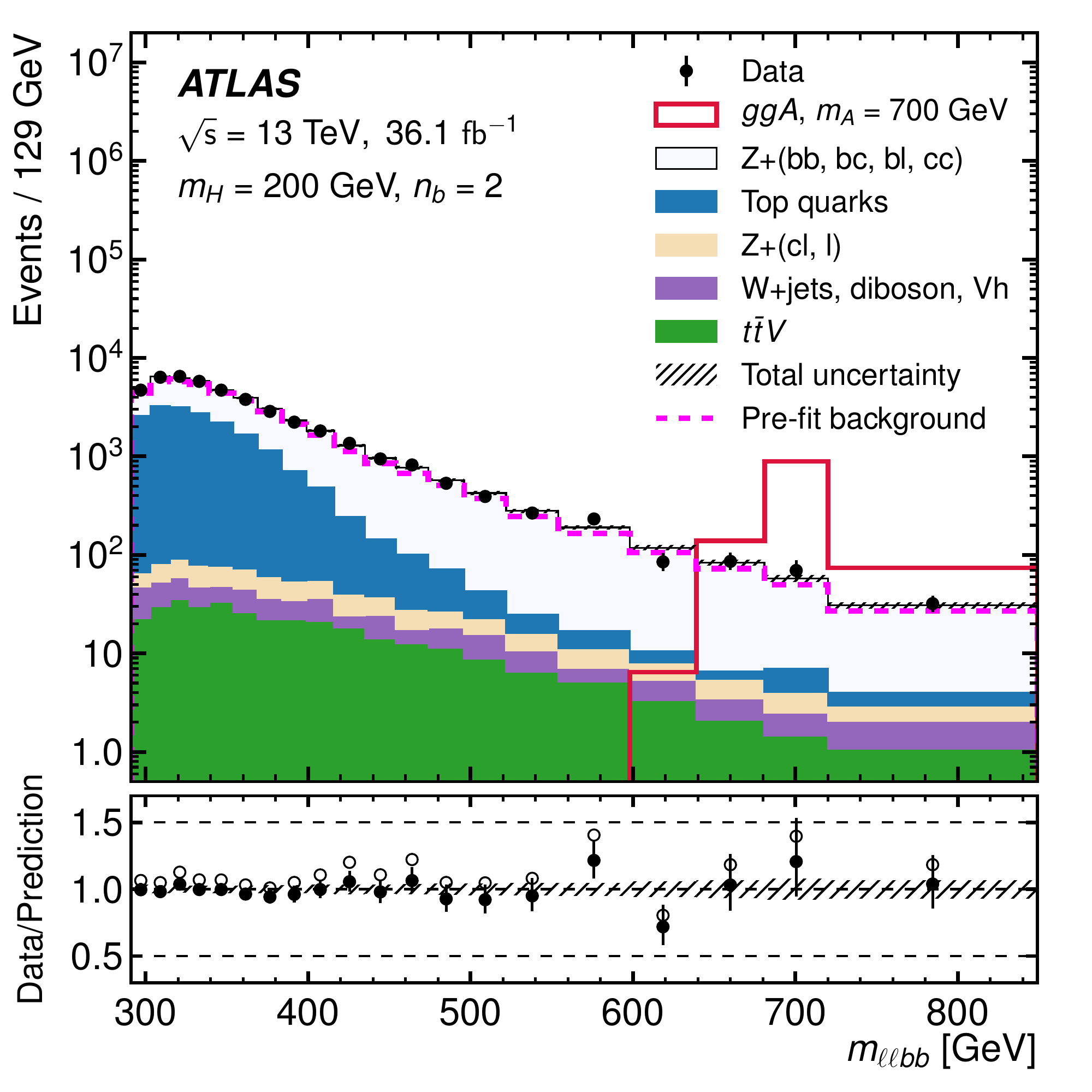}\label{fig:mllbb-2b}}
\subfigure[]{\includegraphics[width=0.49\textwidth]{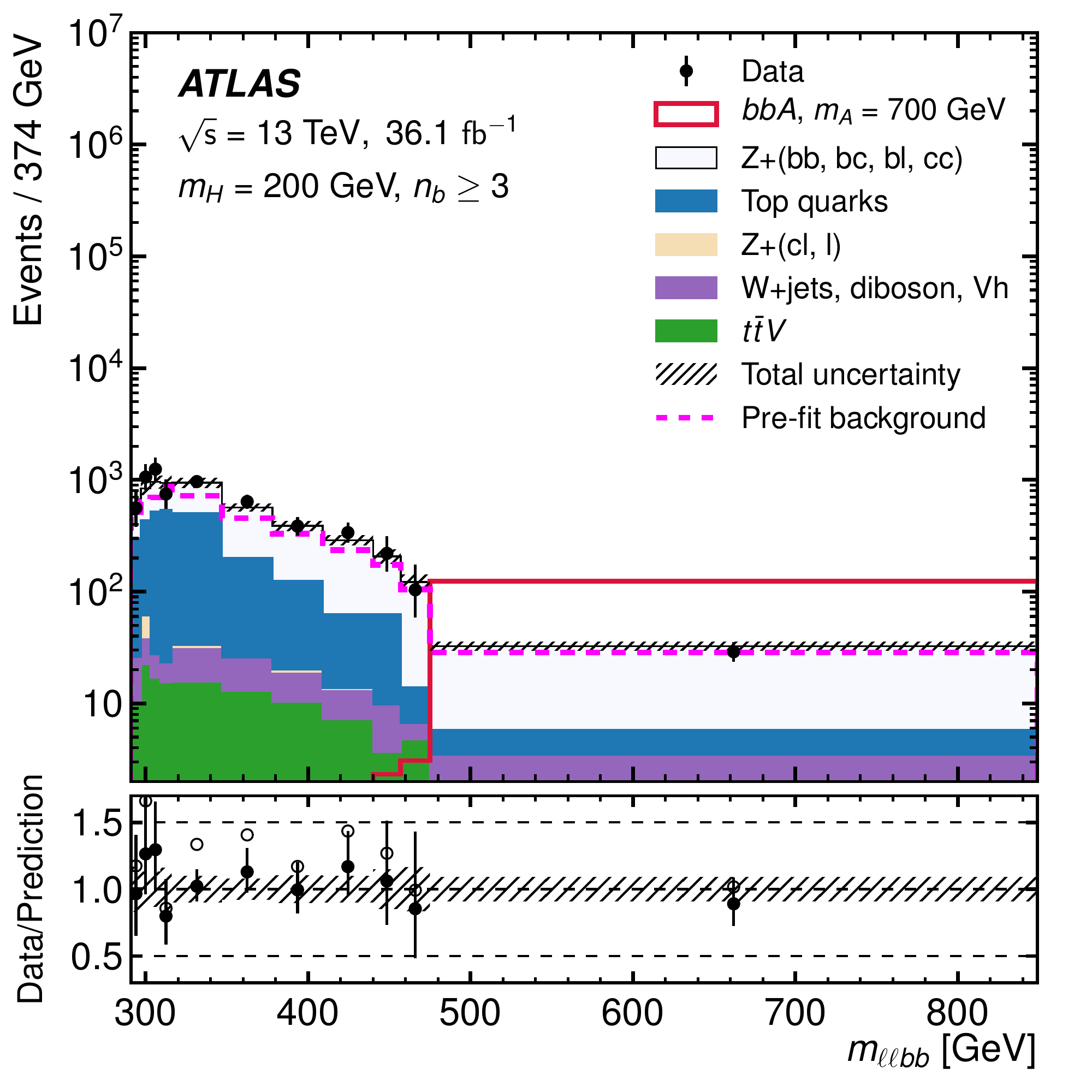}\label{fig:mllbb-3b}} \\
\subfigure[]{\includegraphics[width=0.49\textwidth]{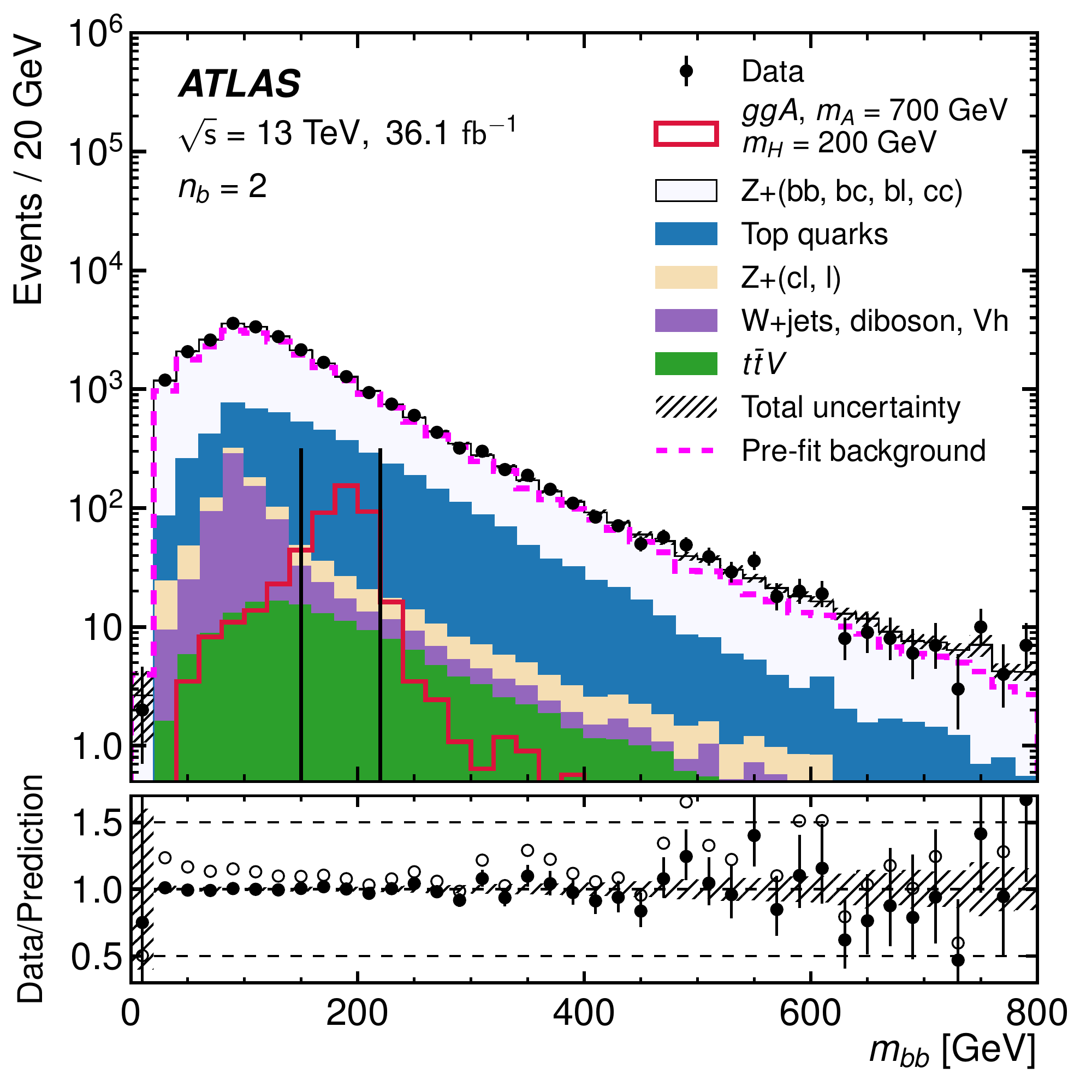}\label{fig:mbb-2b}}
\subfigure[]{\includegraphics[width=0.49\textwidth]{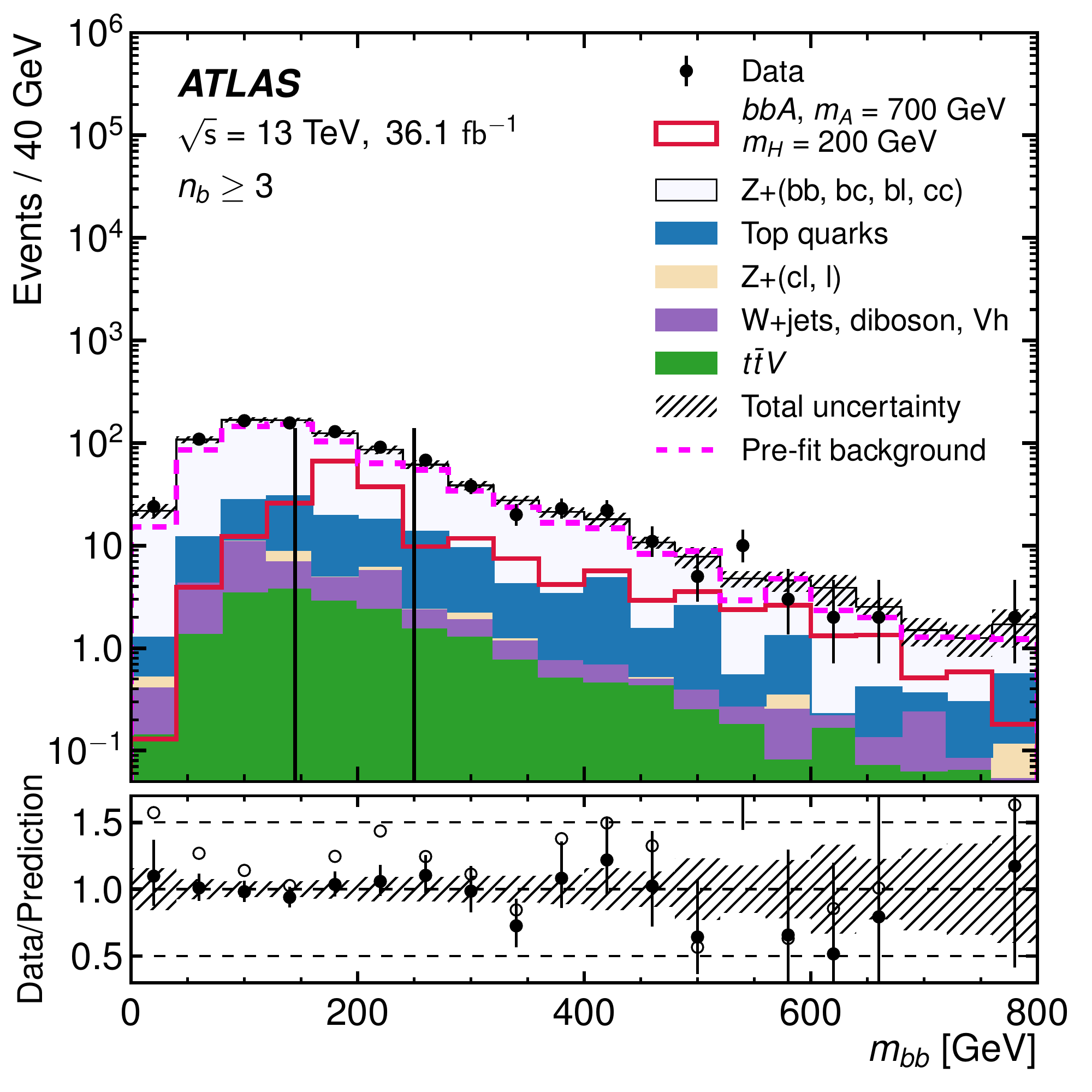}\label{fig:mbb-3b}}
\caption{The $m_{\ell\ell bb}$ mass distribution for the $m_{bb}$ window 
defined for $m_H=200~\GeV$ for (a) the $n_b=2$ and (b) the $n_b\ge 3$ category.
The $m_{bb}$ distribution before any $m_{bb}$ window cuts is shown in 
(c) and (d) for the $n_b=2$ and the $n_b \ge 3$ categories, respectively.
Signal distributions for $m_A=700~\GeV$, $m_H=200~\GeV$ are also shown for 
gluon--gluon fusion production in (a, c) and $b$-associated production in (b, d) assuming
production cross-sections times the branching ratios $\mathcal{B}(A\to ZH)$ and $\mathcal{B}(H\to bb)$ of 1~pb. The solid 
dots in the lower panels represent the ratio of the data to the post-fit 
background prediction, while the open circles are the ratio of the data to the 
pre-fit background prediction.
In the $m_{bb}$ distributions in (c, d) the $m_{bb}$ window boundaries are 
also shown as vertical solid lines. The window efficiencies for the 
signals shown on the plots are about ($77.4\pm 1.0$)\% for (c) and 
($58.3\pm 2.6$)\% for (d). The $m_{\ell\ell bb}$ distributions in (a) and (b) use variable bin widths.
The last bin is used as a reference for normalisation, and its width is noted in the $y$-axis label. In this case, 
the content displayed in a bin is the number of events shown in that bin multiplied by the ratio of widths of 
the last bin relative to the bin shown.  
}
\label{fig:mllbb}
\end{figure}

\begin{figure}[htb]
\centering
\subfigure[]{\includegraphics[width=0.49\textwidth]{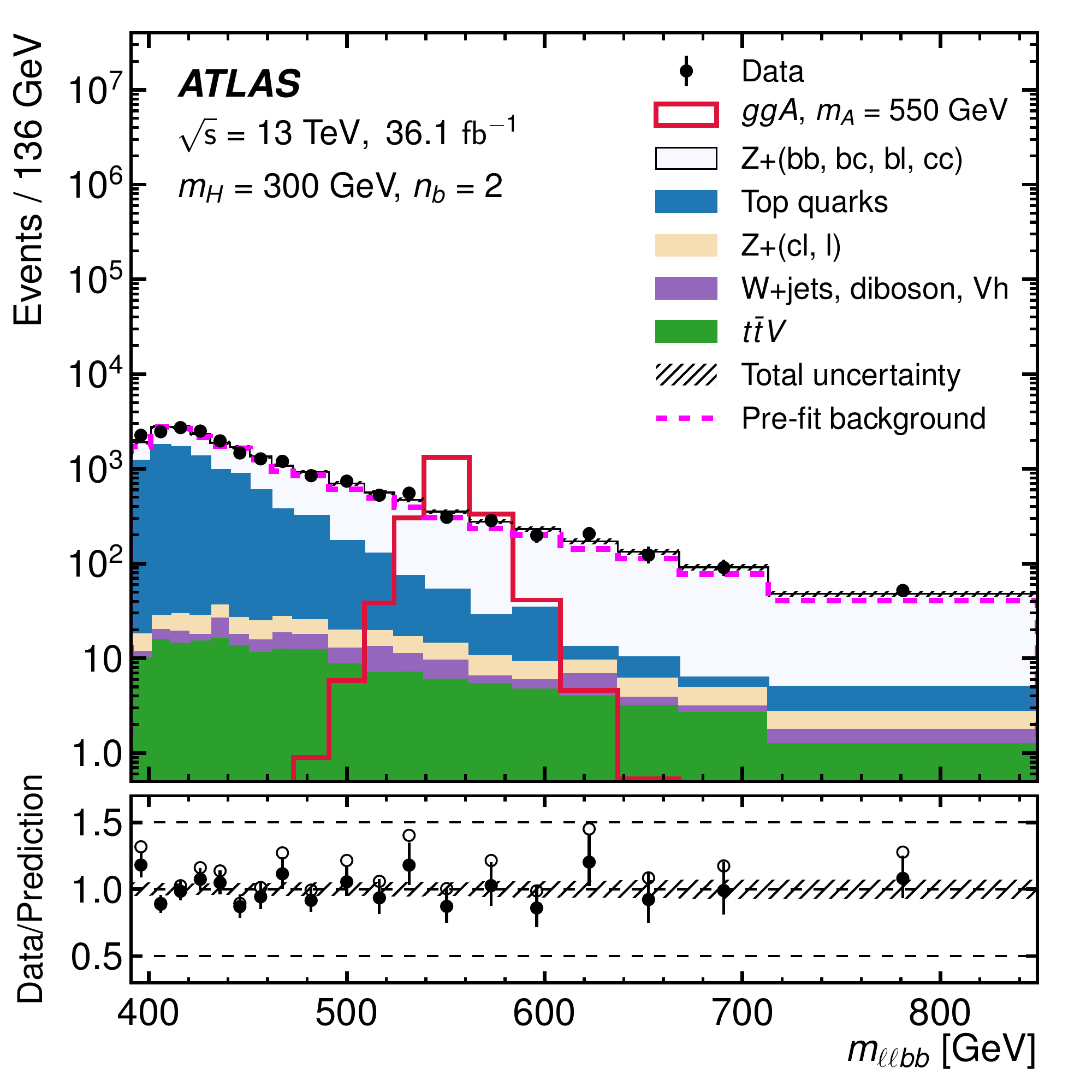}\label{fig:mllbb-1a-2b}}
\subfigure[]{\includegraphics[width=0.49\textwidth]{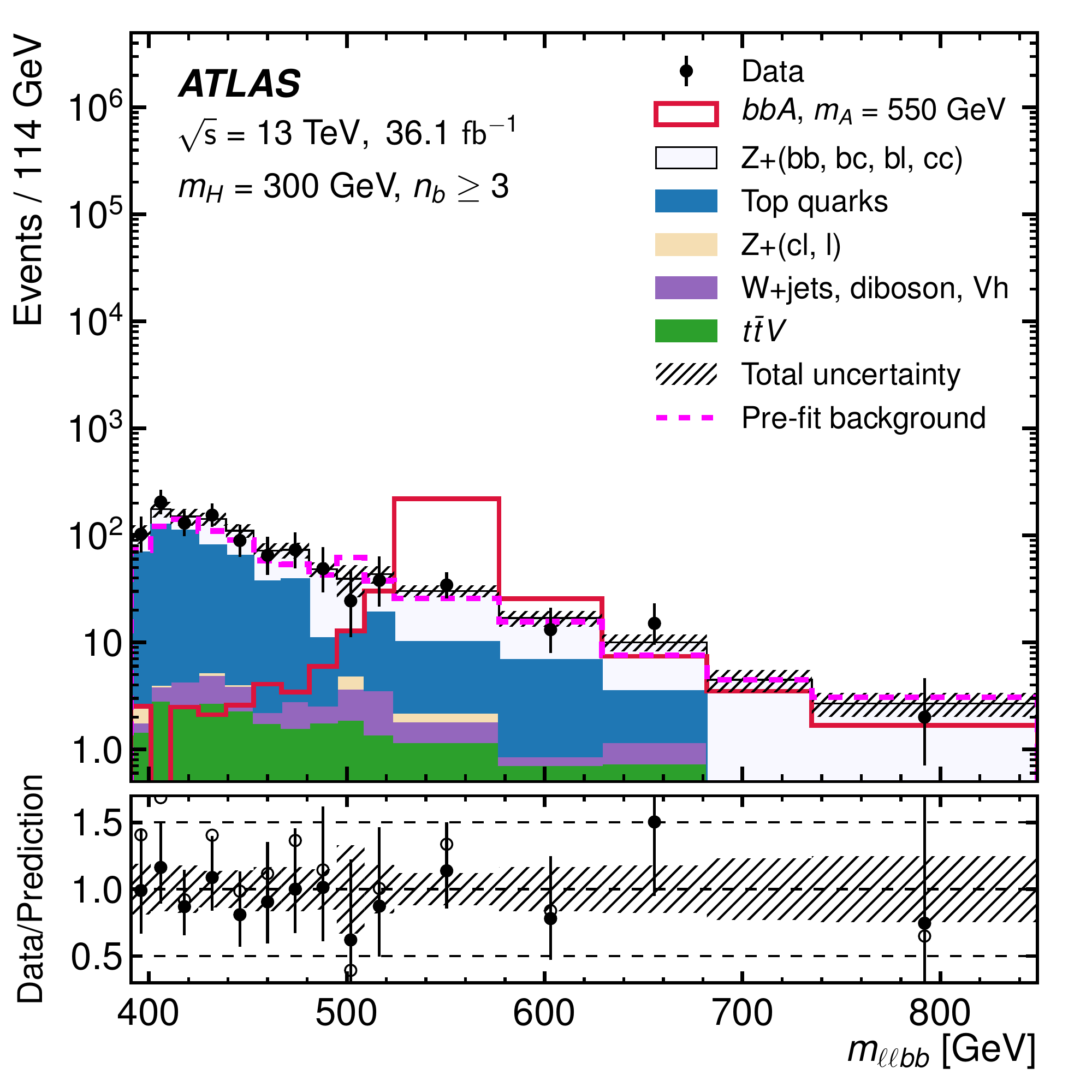}\label{fig:mllbb-1b-3b}} \\
\subfigure[]{\includegraphics[width=0.49\textwidth]{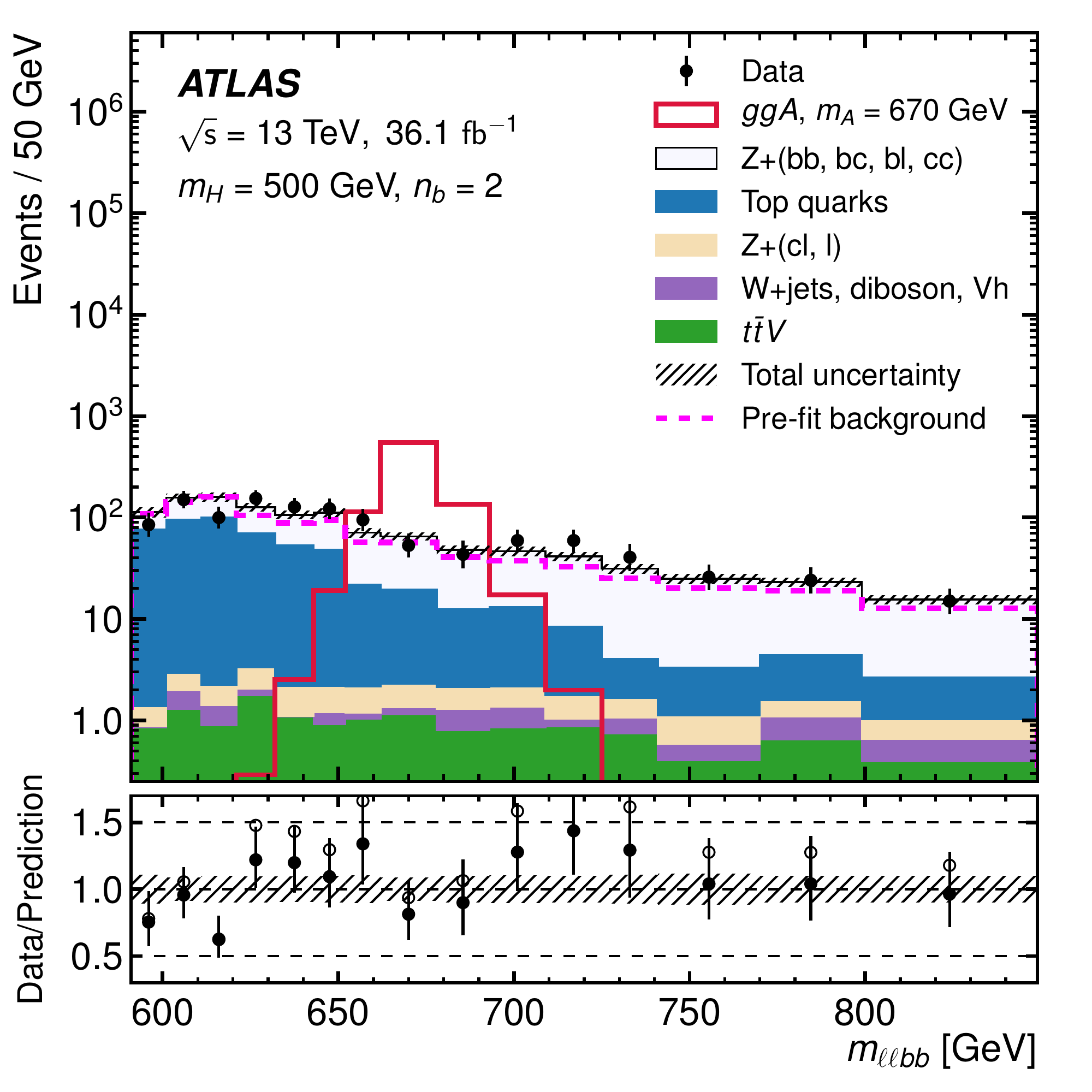}\label{fig:mllbb-1c-2b}}
\subfigure[]{\includegraphics[width=0.49\textwidth]{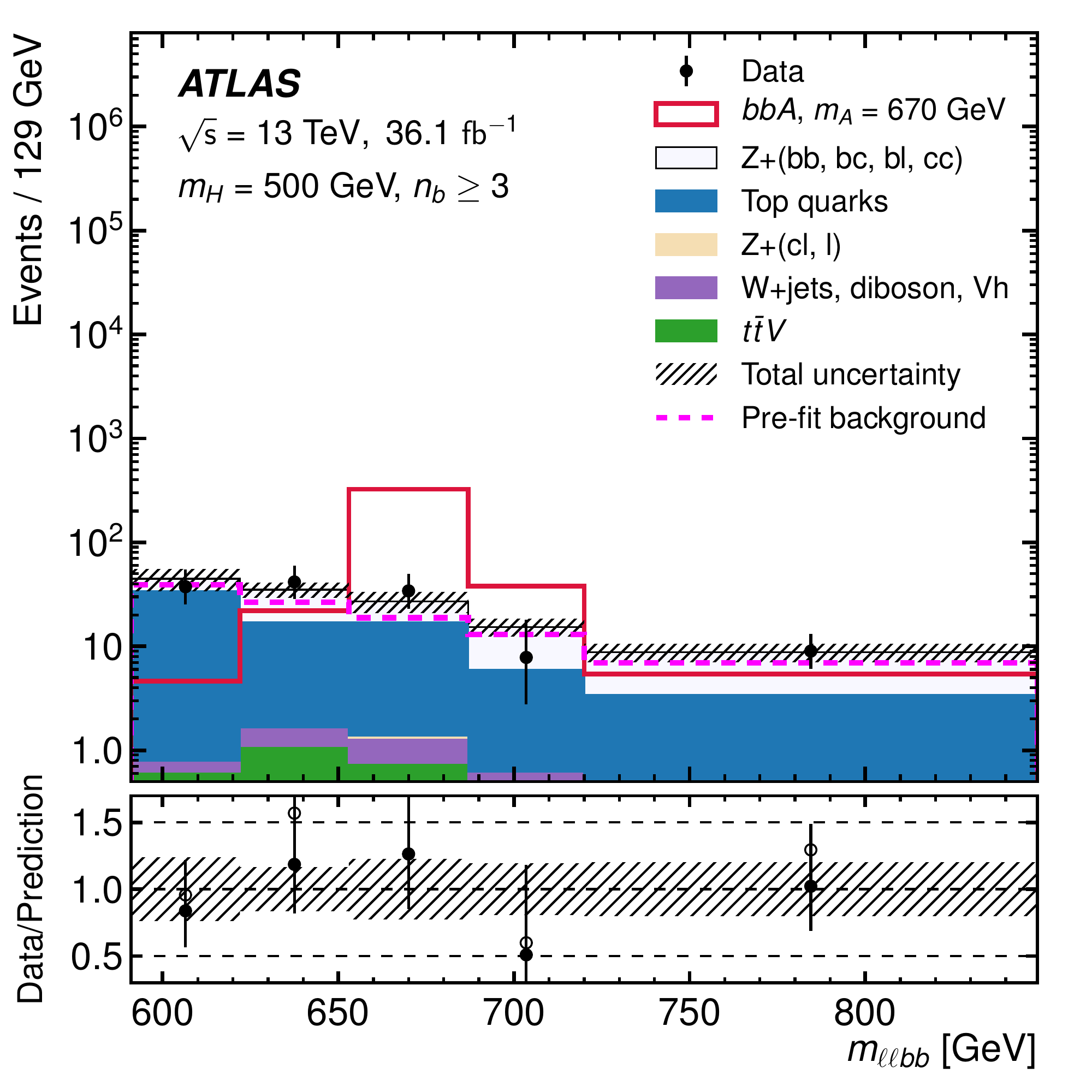}\label{fig:mllbb-1d-3b}}
\caption{The $m_{\ell\ell bb}$ mass distribution for the $m_{bb}$ windows 
defined for $m_H=300~\GeV$ and $m_H=500~\GeV$ for (a, c) the $n_b=2$ and (b, d) the $n_b\ge 3$ category, respectively. Signal distributions are also shown for 
gluon--gluon fusion production in (a, c) and $b$-associated production in (b, d) assuming
production cross-sections times the branching ratios 
$\mathcal{B}(A\to ZH)$ and $\mathcal{B}(H\to bb)$ of 1~pb.
The same conventions as in Figure~\ref{fig:mllbb} are used.}
\label{fig:mllbb-1}
\end{figure}

Figure~\ref{fig:mllbb} shows the $\mllbb$ distributions in the $n_b=2$ 
and $n_b\ge 3$ categories for the $m_{bb}$ window defined for $m_{H}=200~\GeV$.
The $m_{bb}$ distributions before any $m_{bb}$ window cut are also shown in this
figure. The $\mllbb$ distributions for two other $m_{bb}$ windows, defined for
 $m_H=300~\GeV$ and $m_H=500~\GeV$ are shown in Figure~\ref{fig:mllbb-1}.
In all cases, the data are found to be well described by the background model. 
The most significant excess for the gluon--gluon fusion production signal
assumption is at the $(m_A, m_H)=(750,610)$\,\GeV{} signal point, for which the
local (global) significance \cite{Vitells:2011da} is 3.5 (2.0) 
standard deviations. For the $b$-associated production, the most significant
excess is at the $(m_A,m_H)=(510,130)$\,\GeV{} signal point, for which the
local (global) significance is 3.0 (1.2) standard deviations. The significances are
calculated for each production process separately ignoring the contribution from the other.

\begin{figure} 
\centering
\subfigure[]{\includegraphics[width=0.49\textwidth]{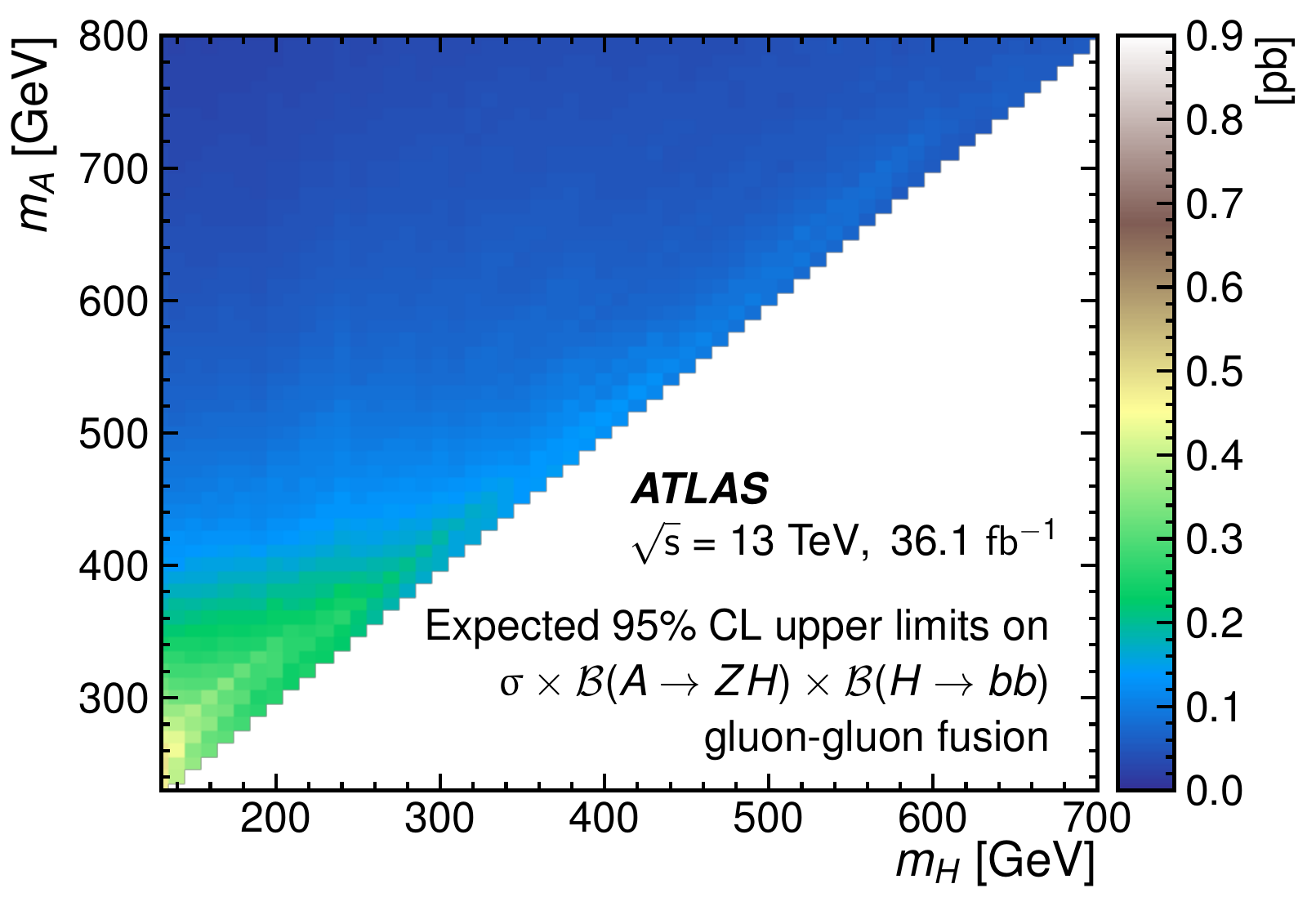}\label{fig:limits:ggA:exp}}
\subfigure[]{\includegraphics[width=0.49\textwidth]{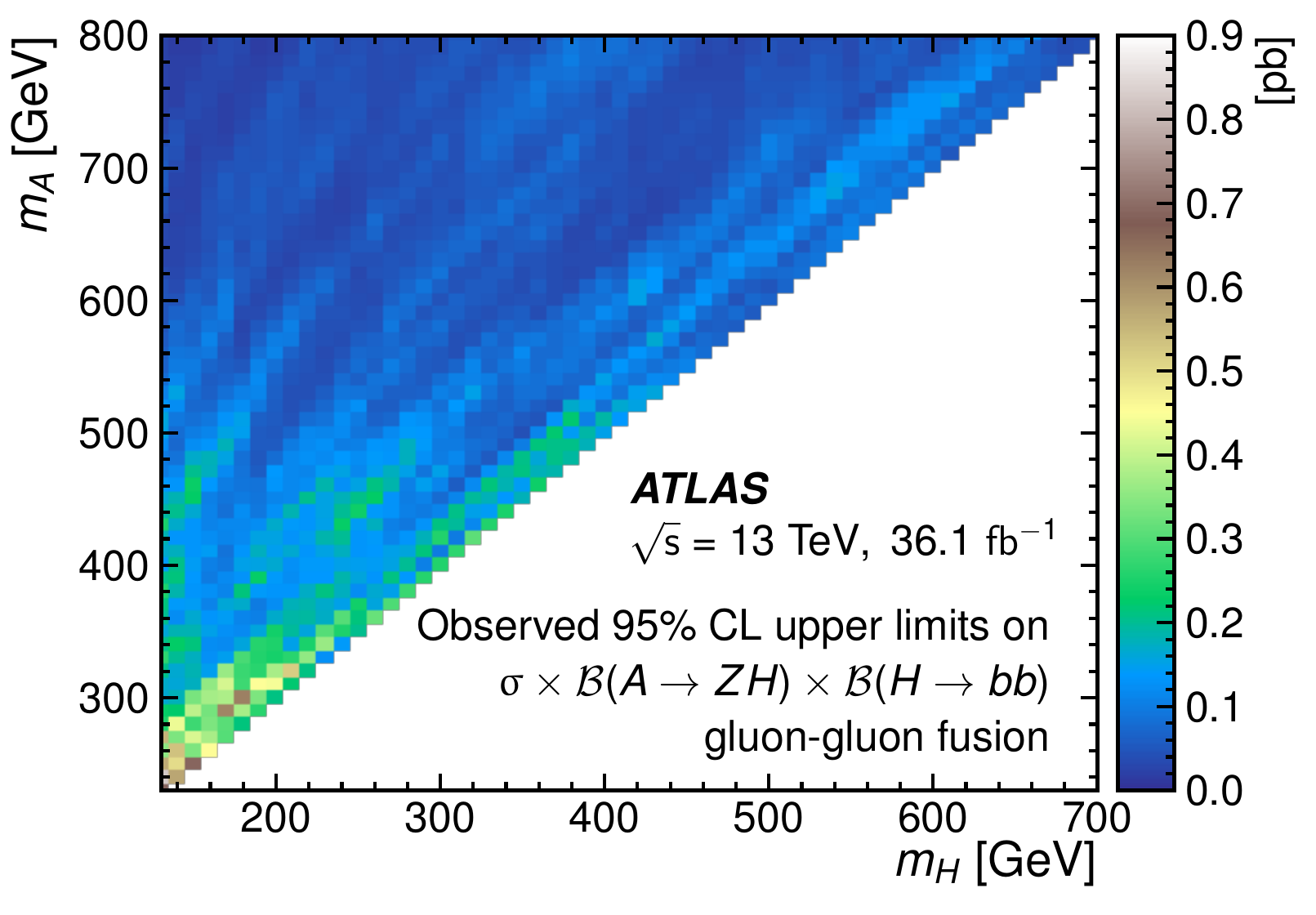}\label{fig:limits:ggA:obs}} \\
\subfigure[]{\includegraphics[width=0.49\textwidth]{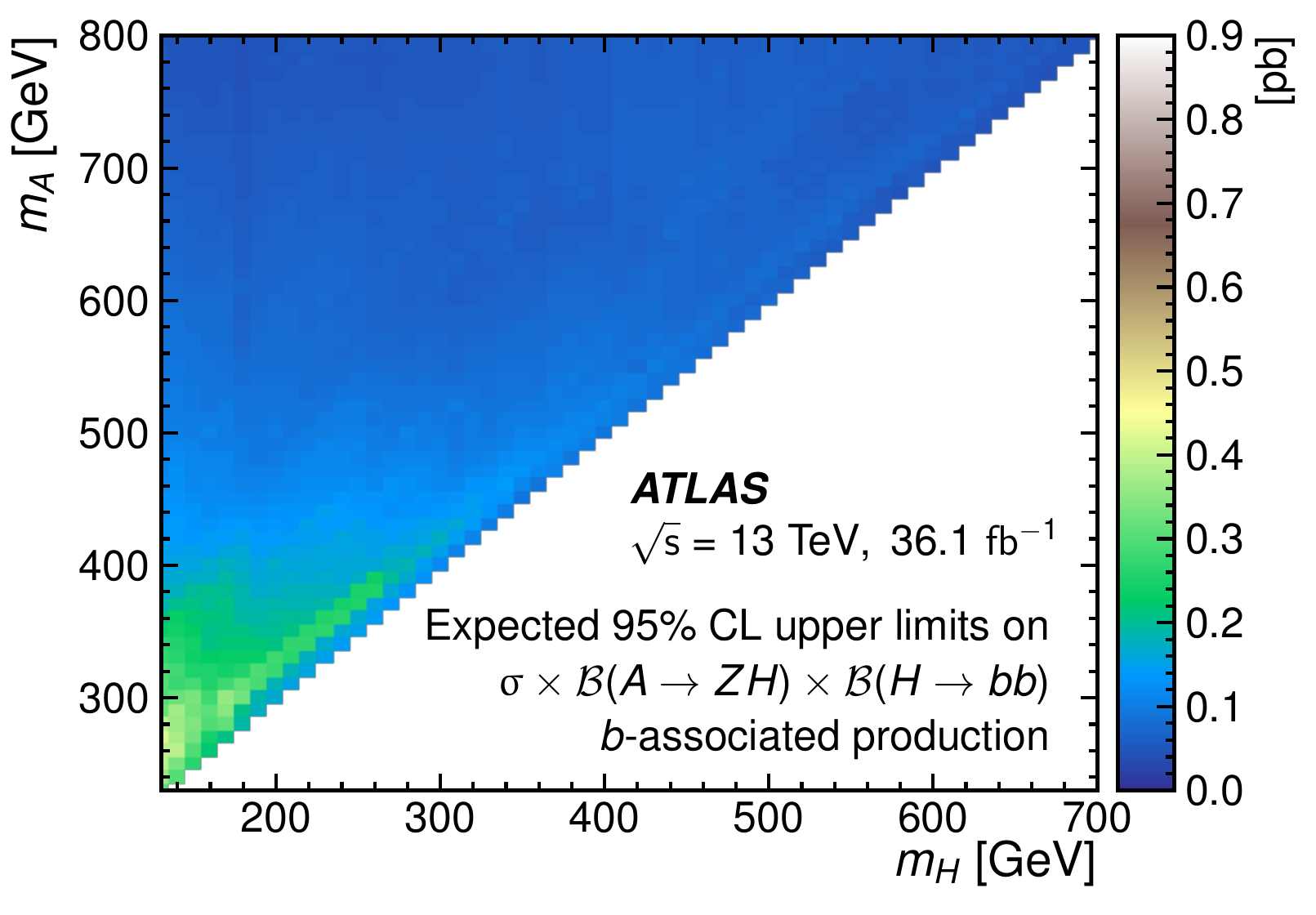}\label{fig:limits:bbA:exp}}
\subfigure[]{\includegraphics[width=0.49\textwidth]{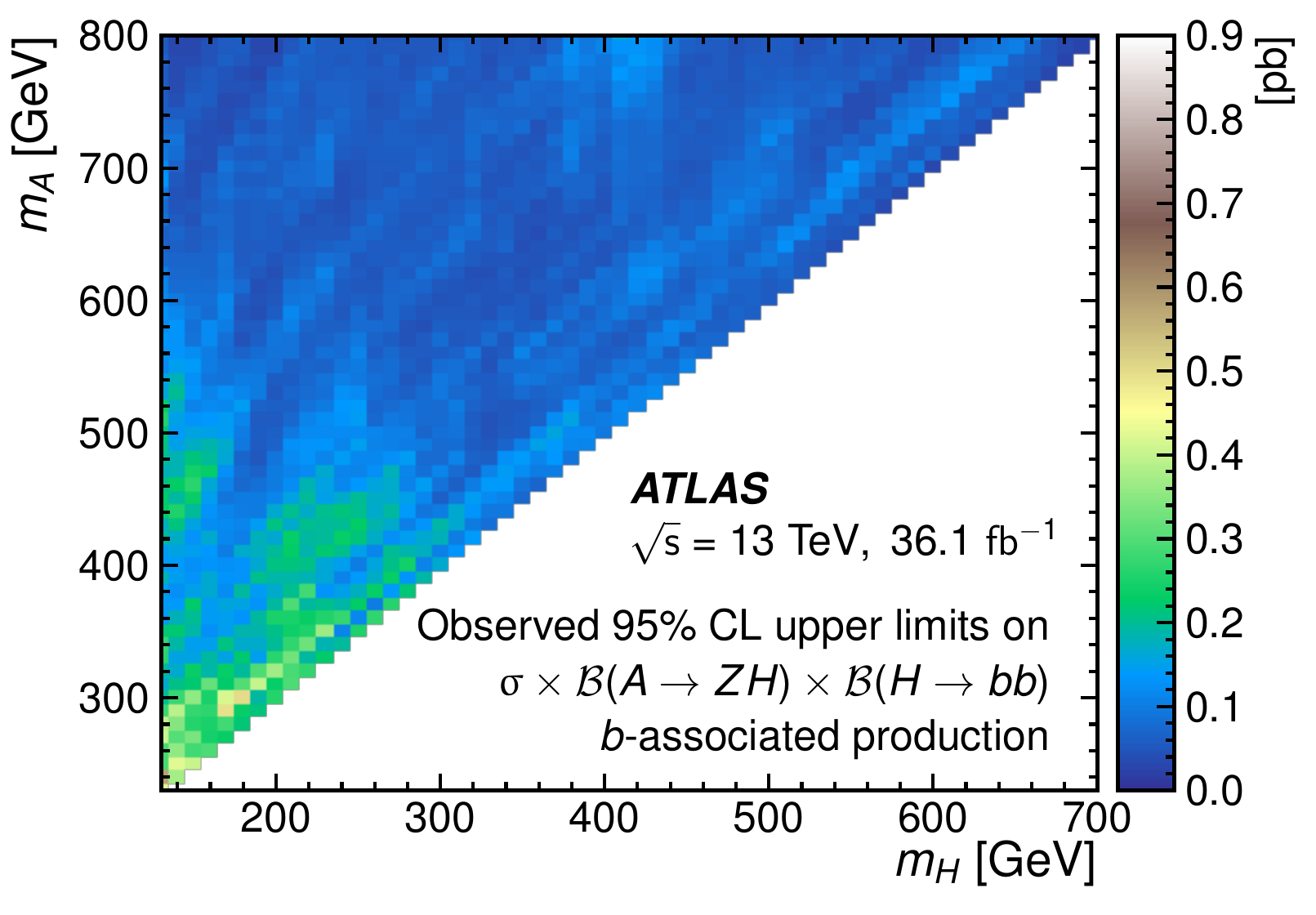}\label{fig:limits:bbA:obs}}

\caption{Upper bounds at 95\% CL on the production cross-section times the branching ratio $\mathcal{B}(A\to ZH)\times \mathcal{B}(H\to bb)$ in pb for (a, b) gluon--gluon fusion and (c, d) $b$-associated production. The expected upper limits are shown in (a) and (c) and the observed upper limits are shown in (b) and (d). 
}
\label{fig:limits}
\end{figure}

In the absence of a statistically significant excess,  constraints on the 
production of $A\to ZH$ followed by the $H\to bb$ decay are derived. 
The method of Ref.~\cite{Read:2002hq} is used to calculate 95\% confidence 
level~(CL) upper bounds on the product of cross-section and decay branching 
ratios, $\sigma\times \mathcal{B}(A\to ZH)\times \mathcal{B}(H\to bb)$, using 
the asymptotic approximation~\cite{Cowan:2010js}. The upper limits are shown in 
Figure~\ref{fig:limits} for a narrow-width $A$ boson produced via gluon--gluon 
fusion and $b$-associated production. As for the significance calculations above, 
these limits are derived separately for each production process. 
For the gluon--gluon fusion limits, 
only the $n_b=2$ category is used. For the $b$-associated production,
both the $n_b=2$ and $n_b\ge 3$ categories are used. The upper 
limit for gluon--gluon fusion varies from \minGGAlim~fb for the 
$(m_A,m_H)=$~\minGGAlimPoint\,\GeV{} signal point to \maxGGAlim~fb for the 
$(m_A,m_H)=$~\maxGGAlimPoint\,\GeV{} signal point. This is to be compared with 
the corresponding expected limits of \minGGAlimExp~fb and \maxGGAlimExp~fb 
for these two signal points. For the $b$-associated production the upper 
limits vary from \minBBAlim~fb for the $(m_A,m_H)=$~\minBBAlimPoint\,\GeV{} 
signal point to \maxGGAlim~fb for the $(m_A,m_H)=$~\maxBBAlimPoint\,\GeV{} 
signal point with expected limits of \minBBAlimExp~fb 
and \maxBBAlimExp~fb, respectively.

The results of the search are interpreted in the context of the 2HDM. 
For this interpretation, several assumptions are made to reduce the number of free 
parameters in the model. The charged Higgs boson is assumed to have the 
same mass as the $A$ boson. The 2HDM parameter $m^2_{12}$ is fixed to 
$m^2_{A}\tan\beta/(1+\tan^2\beta)$. The  lightest Higgs boson of the model, 
$h$, is assumed to have a mass of $125~\GeV$ and its couplings are set to be the 
same as those of the SM Higgs boson, by choosing $\cos(\beta-\alpha)=0$. The widths of 
the $A$ and $H$ bosons are taken from the predictions of 
the 2HDM\@. The cross-sections for $A$ boson production in the 2HDM are 
calculated using up to NNLO QCD corrections for gluon--gluon fusion and 
$b$-associated production in the five-flavour scheme as implemented in 
{\tt SusHi}~\cite{Harlander:2012pb,Harlander:2005rq,Harlander:2003ai,Harlander:2002wh}. For $b$-associated production a cross-section in the four-flavour 
scheme is also calculated as described in Refs.~\cite{Dawson:2003kb,Dittmaier:2003ej} and the results are combined with the five-flavour scheme calculation 
following Ref.~\cite{Harlander:2011aa}. The Higgs boson widths and branching 
ratios are calculated using {\tt 2HDMC}~\cite{Eriksson:2009ws}. The procedure for the calculation of the cross-sections 
and branching ratios, as well as for the choice of 2HDM parameters, follows Ref.~\cite{deFlorian:2016spz}.

\begin{figure}
\centering
\subfigure[]{\includegraphics[width=0.49\textwidth]{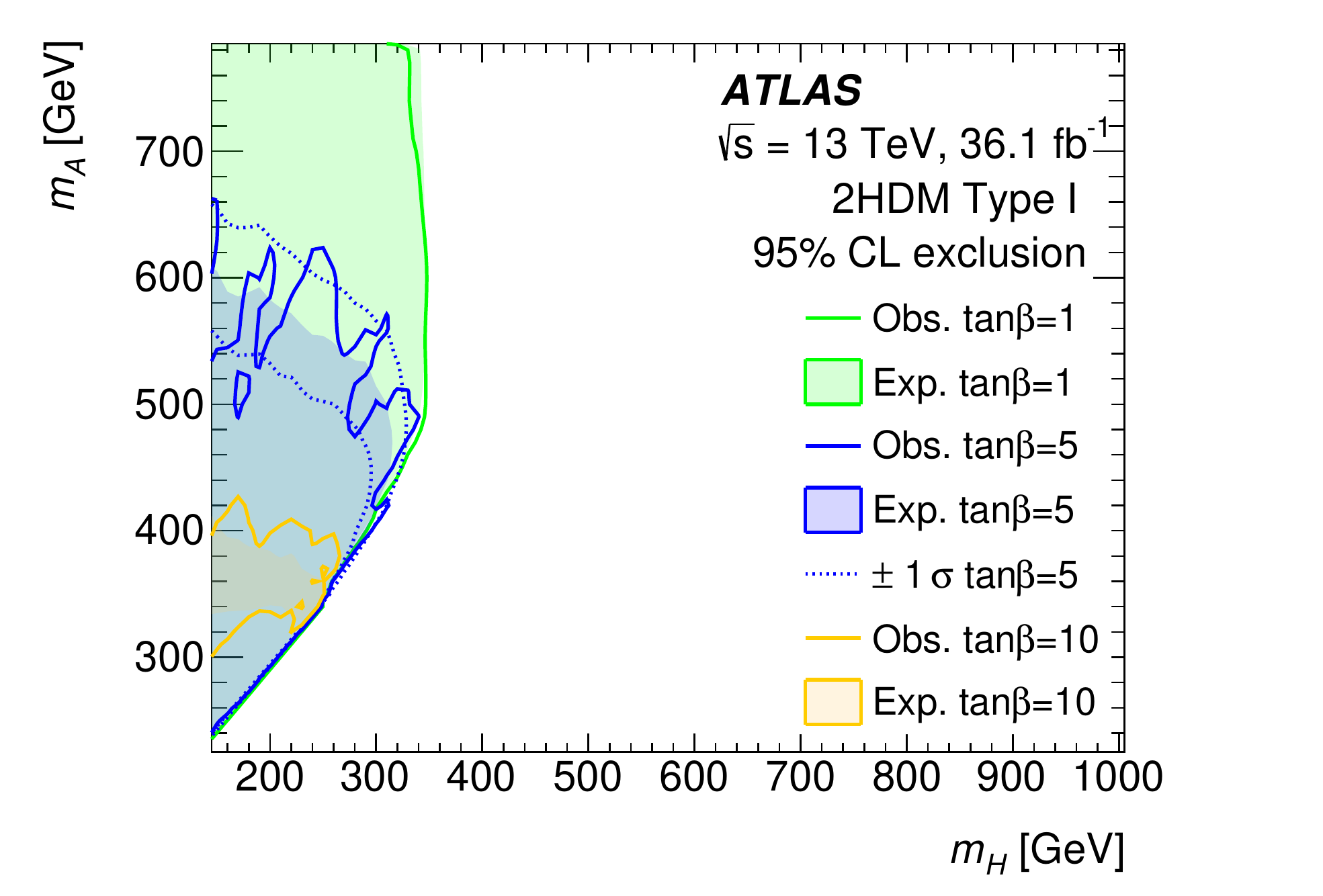}\label{fig:2hdm-type1}}
\subfigure[]{\includegraphics[width=0.49\textwidth]{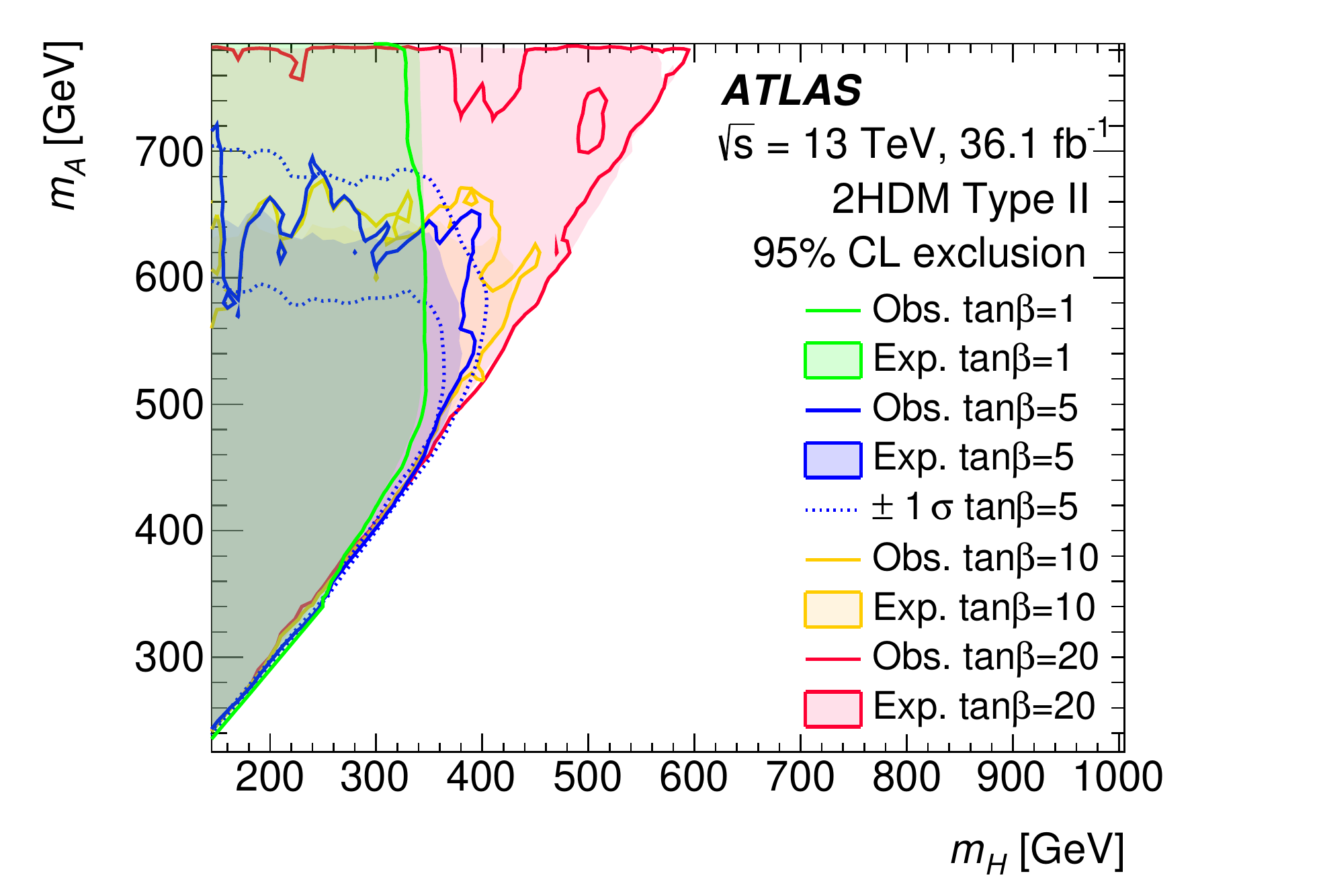}\label{fig:2hdm-type2}} \\
\subfigure[]{\includegraphics[width=0.49\textwidth]{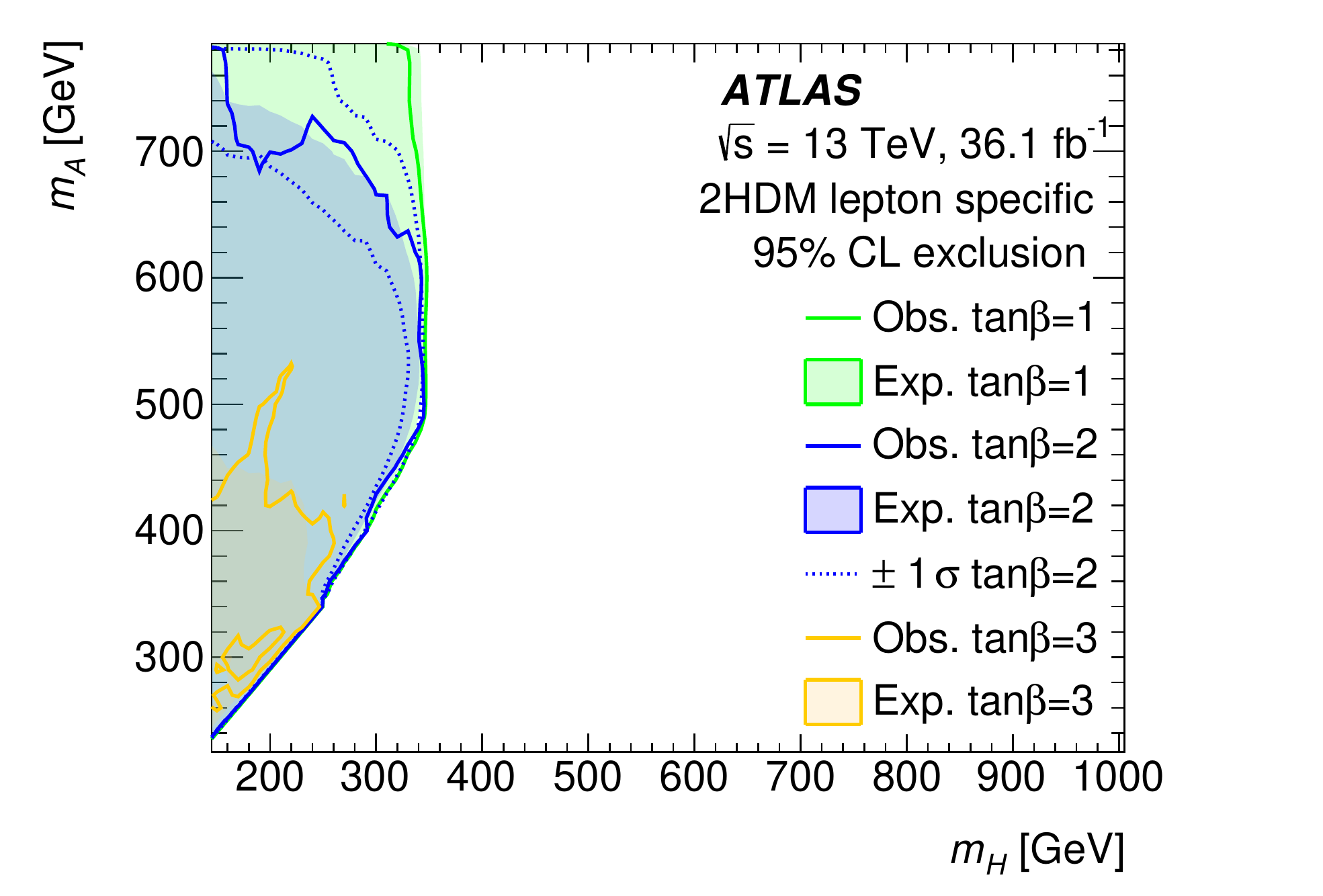}\label{fig:2hdm-type3}}
\subfigure[]{\includegraphics[width=0.49\textwidth]{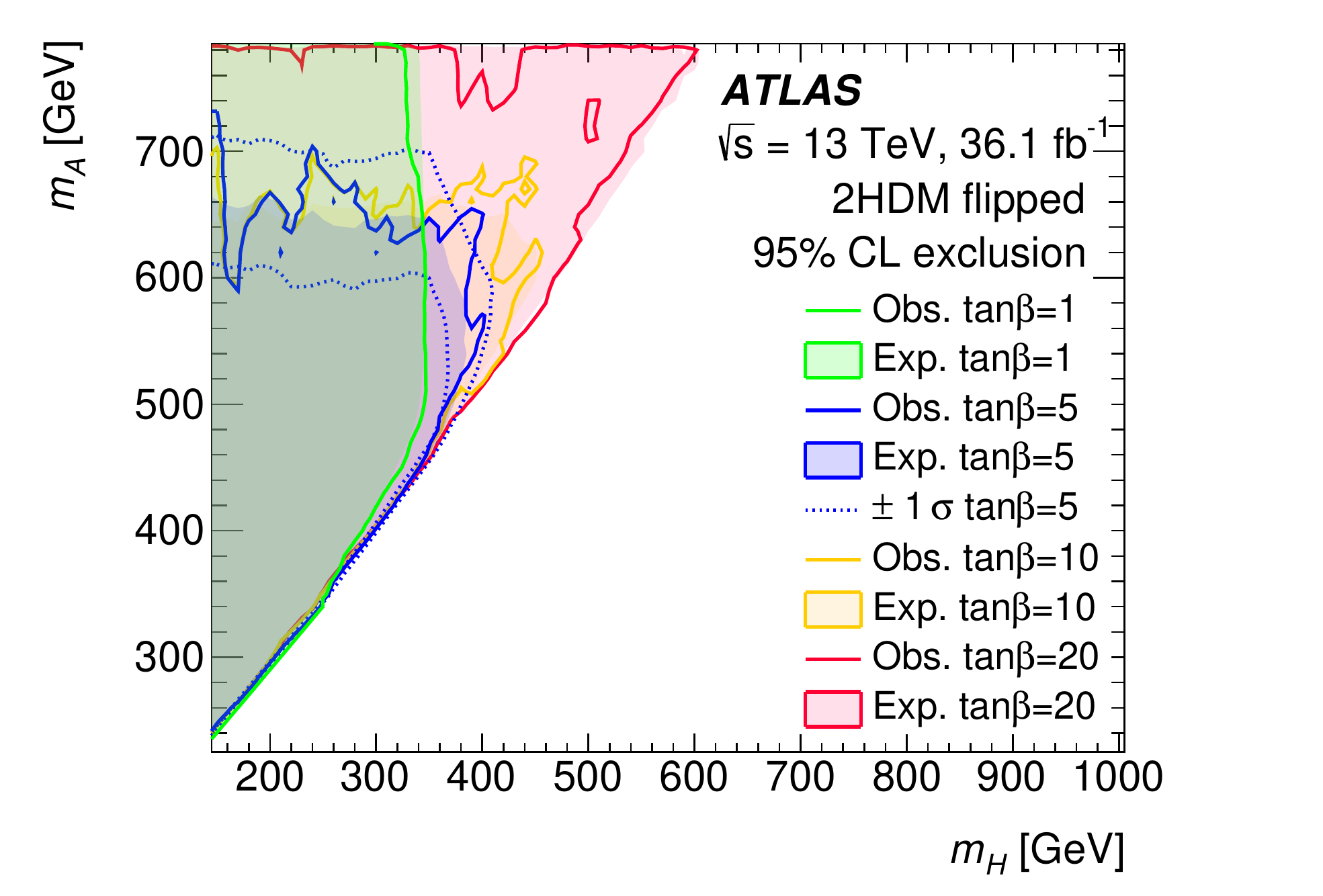}\label{fig:2hdm-type4}}
\caption{Observed and expected 95\% CL exclusion regions in the ($m_A$,$m_H$) plane for various $\tan\beta$ values
for (a) Type I, (b) Type II, (c) lepton specific and (d) flipped 2HDM.}
\label{fig:2hdm}
\end{figure}

Since both gluon--gluon fusion and $b$-associated production are expected, a new signal model weighted by 
the predicted cross sections of the two processes is built for every point tested in the 2HDM parameter space.  
Upper limits on $\sigma\times \mathcal{B}(A\to ZH)\times \mathcal{B}(H\to bb)$ with $\sigma$ here including
contributions from both processes are recalculated and compared
with the 2HDM predictions to derive the limits in the 2HDM parameter space.
Figure~\ref{fig:2hdm} shows the observed and expected limits for Type I, Type II,
`lepton specific' and `flipped' 2HDMs in the ($m_A$, $m_H$) plane for 
various $\tan\beta$ values. Type-II and flipped show similar constraints
because in these models the Yukawa couplings are the same for all fermions apart 
from leptons. The same holds when comparing Type-I and lepton-specific 2HDM,
where the main reason for the difference in sensitivity is the increased significance
of the $H\to\tau\tau$ decay in the lepton-specific model. 
The gluon--gluon fusion production cross section decreases with increasing
$\tan\beta$, which explains the loss of sensitivity in Type-I and lepton specific for
large $\tan\beta$ values. In the case of Type-II and flipped, at large $\tan\beta$ values
it is the $b$-associated production that dominates instead of the gluon--gluon fusion
in this region. For instance the exclusion can reach up to $m_H \approx$~400~\GeV{} 
at lower $\tan\beta$ (less than 10) and $m_H \approx$~600~\GeV{} at higher 
$\tan\beta$ (more than 20).
At low $\tan\beta$ values the Higgs boson branching fraction to $t\bar{t}$
becomes sizable, and this is what limits the sensitivity to below 
$m_H \approx 350$~\GeV\  in all models examined here.

\clearpage

\section{Conclusion}
\label{sec:conclusion}

Data recorded by the ATLAS experiment at the LHC, corresponding to an 
integrated luminosity of \currentluminosity{} from proton--proton collisions 
at a centre-of-mass energy 13~\TeV{}, are used to search for a heavy Higgs 
boson, $A$, decaying into $ZH$, where $H$ denotes a heavy Higgs boson with mass 
$m_H>125~\GeV$. The $A$ boson is assumed to be produced via either gluon--gluon 
fusion or $b$-associated production.  No significant deviation from the SM background 
predictions are observed in the $ZH\to\ell\ell bb$ final state that is 
considered in this search. Considering each production process separately, 
upper limits are set at the 95\% confidence level 
for $\sigma\times \mathcal{B}(A\to ZH)\times \mathcal{B}(H\to bb)$ of 
\minGGAlim--\maxGGAlim~fb for gluon--gluon fusion and \minBBAlim--\maxBBAlim~fb 
for $b$-associated production of a narrow $A$ boson for the mass ranges 130--700 GeV of 
the $H$ boson and 230--800 GeV of the $A$ boson. Taking into account both production processes, 
this search tightens the constraints on the 2HDM in the case of large mass splittings between its 
heavier neutral Higgs bosons.


\FloatBarrier

\section*{Acknowledgements}

We thank CERN for the very successful operation of the LHC, as well as the
support staff from our institutions without whom ATLAS could not be
operated efficiently.

We acknowledge the support of ANPCyT, Argentina; YerPhI, Armenia; ARC, Australia; BMWFW and FWF, Austria; ANAS, Azerbaijan; SSTC, Belarus; CNPq and FAPESP, Brazil; NSERC, NRC and CFI, Canada; CERN; CONICYT, Chile; CAS, MOST and NSFC, China; COLCIENCIAS, Colombia; MSMT CR, MPO CR and VSC CR, Czech Republic; DNRF and DNSRC, Denmark; IN2P3-CNRS, CEA-DRF/IRFU, France; SRNSFG, Georgia; BMBF, HGF, and MPG, Germany; GSRT, Greece; RGC, Hong Kong SAR, China; ISF, I-CORE and Benoziyo Center, Israel; INFN, Italy; MEXT and JSPS, Japan; CNRST, Morocco; NWO, Netherlands; RCN, Norway; MNiSW and NCN, Poland; FCT, Portugal; MNE/IFA, Romania; MES of Russia and NRC KI, Russian Federation; JINR; MESTD, Serbia; MSSR, Slovakia; ARRS and MIZ\v{S}, Slovenia; DST/NRF, South Africa; MINECO, Spain; SRC and Wallenberg Foundation, Sweden; SERI, SNSF and Cantons of Bern and Geneva, Switzerland; MOST, Taiwan; TAEK, Turkey; STFC, United Kingdom; DOE and NSF, United States of America. In addition, individual groups and members have received support from BCKDF, the Canada Council, CANARIE, CRC, Compute Canada, FQRNT, and the Ontario Innovation Trust, Canada; EPLANET, ERC, ERDF, FP7, Horizon 2020 and Marie Sk{\l}odowska-Curie Actions, European Union; Investissements d'Avenir Labex and Idex, ANR, R{\'e}gion Auvergne and Fondation Partager le Savoir, France; DFG and AvH Foundation, Germany; Herakleitos, Thales and Aristeia programmes co-financed by EU-ESF and the Greek NSRF; BSF, GIF and Minerva, Israel; BRF, Norway; CERCA Programme Generalitat de Catalunya, Generalitat Valenciana, Spain; the Royal Society and Leverhulme Trust, United Kingdom.

The crucial computing support from all WLCG partners is acknowledged gratefully, in particular from CERN, the ATLAS Tier-1 facilities at TRIUMF (Canada), NDGF (Denmark, Norway, Sweden), CC-IN2P3 (France), KIT/GridKA (Germany), INFN-CNAF (Italy), NL-T1 (Netherlands), PIC (Spain), ASGC (Taiwan), RAL (UK) and BNL (USA), the Tier-2 facilities worldwide and large non-WLCG resource providers. Major contributors of computing resources are listed in Ref.~\cite{ATL-GEN-PUB-2016-002}.

\clearpage

\printbibliography

\clearpage
 
\begin{flushleft}
{\Large The ATLAS Collaboration}

\bigskip

M.~Aaboud$^\textrm{\scriptsize 34d}$,    
G.~Aad$^\textrm{\scriptsize 99}$,    
B.~Abbott$^\textrm{\scriptsize 124}$,    
O.~Abdinov$^\textrm{\scriptsize 13,*}$,    
B.~Abeloos$^\textrm{\scriptsize 128}$,    
S.H.~Abidi$^\textrm{\scriptsize 165}$,    
O.S.~AbouZeid$^\textrm{\scriptsize 143}$,    
N.L.~Abraham$^\textrm{\scriptsize 153}$,    
H.~Abramowicz$^\textrm{\scriptsize 159}$,    
H.~Abreu$^\textrm{\scriptsize 158}$
% ...
\bigskip
\\

% ...
$^{13}$Institute of Physics, Azerbaijan Academy of Sciences, Baku; Azerbaijan.\\
% ...
$^{34}$$^{(a)}$Facult\'e des Sciences Ain Chock, R\'eseau Universitaire de Physique des Hautes Energies - Universit\'e Hassan II, Casablanca;$^{(b)}$Centre National de l'Energie des Sciences Techniques Nucleaires (CNESTEN), Rabat;$^{(c)}$Facult\'e des Sciences Semlalia, Universit\'e Cadi Ayyad, LPHEA-Marrakech;$^{(d)}$Facult\'e des Sciences, Universit\'e Mohamed Premier and LPTPM, Oujda;$^{(e)}$Facult\'e des sciences, Universit\'e Mohammed V, Rabat; Morocco.\\
% ...
$^{99}$CPPM, Aix-Marseille Universit\'e, CNRS/IN2P3, Marseille; France.\\
% ...
$^{124}$Homer L. Dodge Department of Physics and Astronomy, University of Oklahoma, Norman OK; United States of America.\\
% ...
$^{128}$LAL, Universit\'e Paris-Sud, CNRS/IN2P3, Universit\'e Paris-Saclay, Orsay; France.\\
% ...
$^{143}$Santa Cruz Institute for Particle Physics, University of California Santa Cruz, Santa Cruz CA; United States of America.\\
% ...
$^{153}$Department of Physics and Astronomy, University of Sussex, Brighton; United Kingdom.\\
% ...
$^{158}$Department of Physics, Technion, Israel Institute of Technology, Haifa; Israel.\\
$^{159}$Raymond and Beverly Sackler School of Physics and Astronomy, Tel Aviv University, Tel Aviv; Israel.\\
% ...
$^{165}$Department of Physics, University of Toronto, Toronto ON; Canada.\\
% ...
$^{a}$ Also at  Department of Physics, University of Malaya, Kuala Lumpur; Malaysia.\\
$^{b}$ Also at Borough of Manhattan Community College, City University of New York, NY; United States of America.\\
$^{c}$ Also at Centre for High Performance Computing, CSIR Campus, Rosebank, Cape Town; South Africa.\\
$^{d}$ Also at CERN, Geneva; Switzerland.\\
$^{e}$ Also at CPPM, Aix-Marseille Universit\'e, CNRS/IN2P3, Marseille; France.\\
% ...
$^{*}$ Deceased

\end{flushleft}


\end{document}